# Kinematic Diagnostics of Disks Around Young Stars:
# CO Overtone Emission from WL 16 & 1548C27

JOAN NAJITA[1,2]

Astronomy Department, University of California, Berkeley, CA 94720

JOHN S. CARR[2]

Remote Sensing Division, Naval Research Laboratory
Code 7217, Washington, DC 20375-5351

ALFRED E. GLASSGOLD

Physics Department, New York University
4 Washington Place, New York, NY 10003

FRANK H. SHU

Astronomy Department, University of California, Berkeley, CA 94720

ALAN T. TOKUNAGA[2]

Institute for Astronomy, University of Hawaii
2680 Woodlawn Drive, Honolulu, HI 96822

---

[1], Present address: Harvard-Smithsonian Center for Astrophysics, 60 Garden St., Cambridge, MA 02138

[2] Visiting Astronomer at the Infrared Telescope Facility, which is operated by the University of Hawaii under contract from the National Aeronautics and Space Administration.

**Abstract**

We report high spectral resolution observations of the CO vibrational overtone emission from the young stellar object 1548C27; our observations include both the $v$=2-0 and $v$=5-3 bandhead regions. These data and similar observations of the young stellar object WL16, reported in a previous contribution to this journal (*Letters*), provide some of the most compelling evidence to date for the existence of inner disks around young stars. We describe the simple procedure that we use to synthesize bandhead emission from disks including the effect of thermal dissociation of CO and non-LTE excitation of the vibrational levels. Using this spectral synthesis procedure to extract the kinematics and physical properties of the emitting gas from the overtone data, we show how these high signal-to-noise data are also powerful probes of the stellar and inner disk properties of these systems. Our modeling is consistent with the identification of WL 16 and 1548C27 as Herbig AeBe stars with stellar masses of approximately 2 and 4 $M_\odot$, respectively. Thus, the kinematic signature of rotating disks in the overtone spectra of these sources provide strong support for the role of accretion disks in the formation of *intermediate mass* stars. For both WL 16 and 1548C27, we interpret our modeling results as indicating that the overtone emission arises from a temperature inversion region in the inner disk atmosphere. We also find evidence for suprathermally broadened lines and are able to place useful constraints on the radial temperature and column density distributions of the CO line-formation region of the disk atmosphere. Given these deduced properties, we discuss the constraints that our observations place on the physical processes responsible for the overtone emission in these sources.

*Subject headings:* accretion, accretion disks — infrared: stars — line: profiles — line: formation — stars: circumstellar matter — stars: formation — stars: pre-main sequence

## 1. Introduction

Circumstellar disks play a fundamental role in the formation of stars and planets. Stars are believed to build up a substantial fraction of their main-sequence mass by accretion through circumstellar disks and these disks constitute the reservoirs of mass for the formation of both planets and close stellar companions. Various pieces of observational evidence assembled over the past decade have been used to argue for the presence of accretion disks around young stars. These include the spectral energy distributions of young stars (e.g., Adams, Lada, & Shu 1987), the asymmetry of their optical forbidden line profiles (e.g., Edwards et al. 1987), their millimeter and submillimeter line and continuum emission (cf. Ohashi & Hayashi 1995; Koerner & Sargent 1995 and references therein; Keene & Masson 1990; Lay et al. 1994), and the cross-correlation linewidths of optical and near-infrared lines of FU Ori objects (e.g., Hartmann & Kenyon 1987a,b). More recently, direct *kinematic* evidence for the existence of circumstellar disks has been obtained from high spectral resolution observations of the CO overtone emission of young stars (Carr et al. 1993, Chandler et al. 1993, 1995). Since the CO overtone lines arise from the inner regions of these disks, detailed studies of this emission can be used to determine the properties of disks within several tens of stellar radii of young stars.

This inner disk region ($< 1$ AU) is of particular interest as it is believed to be dynamically active, giving rise to energetic inflows and outflows of gas which remove substantial amounts of mass and angular momentum from the system and regulate the rotation periods of young stars (Shu et al. 1994; cf. Edwards et al. 1993; and Bouvier et al. 1993). Although regions of this size cannot be imaged directly with present observational techniques, they can be studied with high resolution near-infrared spectroscopy of lines that selectively probe the physical conditions of the region. The exceptional stability of the CO molecule makes it a versatile probe of this region and the excitation temperatures and critical densities of the overtone bands near $2.35\mu$m are particularly well suited to the study of the high densities (up to $10^{15}$ cm$^{-3}$) and temperatures (1000–6000K) that characterize the inner regions of young stellar objects (YSOs), e.g., as first demonstrated by Scoville et al. (1983) for the BN object.

With the advent of a new generation of sensitive near-infrared spectrometers (e.g., CSHELL, Tokunaga et al. 1990), it is now possible to study *low-* and *intermediate-mass* YSOs in the near-infrared at high spectral resolution. Using these spectrometers, even faint, low-luminosity ($\sim 10 L_\odot$) YSOs can now be observed with a resolution up to 40,000 and at a signal-to-noise that was previously only feasible for bright, high-luminosity sources (e.g., Mitchell et al. 1989, 1990) and FU Ori objects (Hartmann & Kenyon 1987a,b). We have previously reported spectroscopic observations of the $v$=2-0 bandhead region of the embedded sources SVS 13 (Carr & Tokunaga 1992) and WL 16 (Carr et al. 1993). Similar observations have been reported by Chandler et al. (1993) for several additional sources. Our recent observations of the $v$=2-0 and $v$=5-3 bandhead regions of the Herbig AeBe star 1548C27 are presented here. A striking aspect of these studies is that the bandhead often shows the characteristic shape of emission from a rotating disk. The WL 16 and 1548C27 spectra provide particularly good examples and clearly demonstrate the existence of inner disks in these systems.

Winds from low-luminosity YSOs may also produce substantial CO overtone emission (Carr 1989; Chandler et al. 1995; Safier et al. 1995). Calculations of the thermal and chemical structure of winds from these sources show that CO is a robust chemical product and that the inner wind region is warm enough to excite overtone emission ($\sim$ 2000K; Ruden, Glassgold, & Shu 1990). Since the densities of winds rapidly drop below the critical density for the overtone bands (cf. Appendix B), a significant wind contribution to the overtone emission is expected to be restricted to YSOs with large mass loss rates. It is less clear whether a large CO abundance can be expected to survive possible photodissociation by the stellar radiation field of the earlier stellar spectral type sources in the present study. In any case, overtone emission from winds is expected to show a line asymmetry due to the occultation of the receding flow by optically thick inner disks. Since the wind is likely to be cooler than the star and to have a temperature similar to that of the inner disk region (Ruden, Glassgold, & Shu 1990; Chandler et al. 1995), the absorption of stellar and inner disk continuum photons by the wind will introduce an additional line asymmetry. The absence of these asymmetries in the overtone spectra in the present study suggests that the emission from these sources is not dominated by the contribution from a wind.

In this paper, we demonstrate that our overtone observations of WL16 and 1548C27 are well-explained by a disk interpretation for the emission. We develop and implement a simple spectral synthesis technique in order to illustrate the constraints that these high resolution, high signal-to-noise data place on both basic stellar properties (e.g., masses and radii) as well as detailed disk properties (e.g., radial variations of temperature, column density, and intrinsic linewidth). In previous publications, we used a spectral synthesis procedure developed by Carr et al. (1993), which generalized previous work addressing isolated lines (e.g., Huang 1972, Smak 1984) to the bandhead. The synthesis problem was formulated in terms of an effective emissivity as a function of disk radius; inner and outer disk radii were introduced as fitting parameters. In addition, the line emitting region was assumed to correspond to the entire vertical disk column density and to be optically thin in the continuum (see also Carr 1989, Chandler et al. 1995).

In the present application, we develop a more general synthesis method formulated in terms of the physical properties of the CO line-formation region of the disk atmosphere. We use a two-layer approximation to this atmospheric region: the continuum forms in the lower layer and the CO lines form in the upper layer. Thus, the two-layer model approximates the situation of an optically thick disk with a temperature inversion in the upper disk atmosphere (cf. Calvet et al. 1991). We adopt radial power-law distributions of temperature in both layers and of column density in the upper layer. Our analysis then yields the radial variation of physical conditions that must be explained by a more fundamental future theory. We avoid the introduction of arbitrary inner and outer disk radii by explicitly considering the abundance and excitation of CO. In addition to placing tight constraints on the temperature and column density distributions of the emitting layer, we find that the intrinsic (local) linewidth of the emitting gas is significantly broader than thermal Doppler values, probably symptomatic of supersonic MHD turbulence.

We begin, in §2, by defining the two-layer disk atmosphere model and outlining the procedure used to synthesize the CO spectrum of a rotating disk from known level populations. In the next three sections, we discuss the applicability of LTE level populations and the circumstances under which departures from LTE become important. In §3, we model the $v$=2-0 bandhead spectrum of WL 16 assuming LTE in order to illustrate the parameter dependence of the two-layer model and extract the basic physical properties of the emitting gas. In §4, we develop a non-LTE theory for the level populations and draw attention to the diagnostic power of jointly fitting both the $v$=2-0 and $v$=5-3 bandhead regions. The utility of modeling these two spectral regions is demonstrated in §5 where we present our observations of the Herbig AeBe star 1548C27 in these spectral regions and use our spectral synthesis procedure to deduce the physical properties of the CO line formation region in this source. Given these deduced properties, in §6 we discuss the evolutionary status of WL 16 and 1548C27 and comment on the physical processes that are likely to be responsible for the CO emission in these sources.

## 2. A Two-Layer Disk Atmosphere Model

We model the CO overtone emission as arising from the inner ($\lesssim$ 1AU) region of a geometrically thin disk in Keplerian rotation about a star of mass $M_*$. We treat the atmosphere as two adjacent layers, each of which is vertically homogeneous. The overtone lines form in the upper ("line") layer, which extends vertically over a thermal scale height and has negligible continuum opacity. The latter assumption is checked *a posteriori* by computing the H$^-$ continuous opacity. There is no scattering in the upper layer and the only line opacities are those of the CO lines themselves. The lower ("continuum") layer is optically thick in both the overtone lines and in the 2.3$\mu$m continuum.

We adopt power-law radial temperature distributions for the continuum and line layers: $T_c(r) = T_{c0}(r/r_0)^{-p_c}$ and $T_\ell(r) = T_{\ell 0}(r/r_0)^{-p_\ell}$ where $r_0$ is a fiducial disk radius. The line layer is additionally described by a vertical mass column density distribution that also varies as a power-law with radius: $\sigma_\ell(r) = \sigma_{\ell 0}(r/r_0)^{-q_\ell}$. We assume solar abundances of gaseous carbon and oxygen in calculating the chemical equilibrium abundance of CO in this layer. The additional parameters required to synthesize the absolute strength and shape of the spectrum near the CO bandheads are: the local line profile function ($\phi_\nu$), the distance and $K$-band extinction to the system ($d$, $A_K$), the inclination of the rotation axis of the disk to the line of sight ($i$), and the stellar luminosity and radius ($L_*$, $R_*$). The last two quantities determine the stellar contribution to the 2.3 $\mu$m continuum which is assumed to be blackbody.[1] Of these, the stellar and system parameters—$M_*$, $L_*$, $R_*$, $d$, $A_K$, and $i$—are relatively well constrained by the gross properties of the CO overtone emission and other observational data (see, e.g., Carr et al. 1993). The disk properties—$T_c$, $T_\ell$, $\sigma_\ell$, and $\phi_\nu$—which are the quantities of primary interest here, are well constrained by the type of detailed modeling that we discuss below.

The synthesis procedure allows the introduction of inner and outer disk truncation radii ($R_{\rm in}$, $R_{\rm out}$) as model parameters and explicitly considers the variation of CO abundance and excitation with disk radius. Both disk truncation and these latter processes may limit the radial extent of the CO emission. For example, disks may be physically truncated from the inside at distances of several stellar radii by stellar magnetospheres, as suggested by a variety of recent observational and theoretical results (c.f. Bertout et al. 1988, Königl 1991, Bouvier et al. 1993, Edwards et al. 1993, Hartmann et al. 1994, Shu et al. 1994). In addition, the dissociation of CO at temperatures $T_\ell \gtrsim 5000$ K will naturally introduce an effective inner radius to the CO emission; in contrast to the case of physical disk truncation, contributions to the 2.3$\mu$m continuum may arise from within the CO dissociation radius. Disks may also be physically truncated from the outside by the formation of giant planets or the presence of nearby stellar companions (e.g., Lin & Papaloizou 1993, Mathieu 1994). Since the YSOs in the present study are not close binary systems and giant planets are believed to form outside the water-ice condensation radius (at $T \sim 150$ K)—well beyond the region responsible for the CO overtone emission—we generally set the outer radius of the disk $R_{\rm out}$ at infinity. In this case, effective outer radii for the CO emission are produced by either dust formation or departures from LTE populations (see §§4 and 6).

To synthesize the bandhead emission, the disk surface is first divided radially and azimuthally into finite elements. For each element, we calculate the emergent intensity

$$I_\nu = B_\nu(T_c)e^{-\tau_\nu/\mu} + S_\nu(1 - e^{-\tau_\nu/\mu}), \qquad (1)$$

and then add a frequency shift appropriate to the projected rotational velocity of the element. In the equation above, $B_\nu(T_c)$ is the Planck function at the temperature of the continuum layer, $\mu = \cos i$, and $S_\nu$ and $\tau_\nu$ are the source function and total vertical optical depth through the upper layer of the element. Both the source function and optical depth of the upper layer may include contributions from discrete lines that overlap in frequency. The line layer is assumed to be spatially thin so that the properties of the disk atmosphere are approximately constant along the inclined ray path. Having calculated the intensity from each element, the contributions of all elements not shadowed by the star are then summed to obtain the disk contribution to the spectrum. Finally, we add the stellar (blackbody continuum) contribution to the spectrum, accounting for the shadowing of the star by the disk. The resulting spectrum is extincted by the adopted value of $A_K$ to produce the final synthesized spectrum.

---

[1] We ignore the stellar contribution to the CO overtone lines. While this approximation is suitable for the probable early-type sources discussed in this paper, the contribution of the stellar absorption lines to the composite YSO spectrum is required to model later-type sources such as DG Tau (cf. Chandler et al. 1993).

## 3. LTE Model

In order to illustrate the dependence of the bandhead emission on the parameters of the two-layer model, in this section we apply the above spectral synthesis procedure to our $v$=2-0 data for WL 16 assuming LTE level populations. The rotational and vibrational constants, and the energy levels are taken from Mantz et al. (1975). The use of LTE level populations allows us to extract the basic kinematics of the emitting gas from the bandhead spectrum in a simple way, i.e., without having to address the additional complication of non-LTE effects which generally bear weakly on the deduced kinematics. The LTE results also form a basis of comparison against which we can measure the magnitude of the non-LTE corrections that are computed in §§4 and 5.

### 3.1. Spectral Synthesis of the v=2-0 Bandhead Region

The spectral region covered by our observations of the $v$=2-0 bandhead emission from WL 16 (cf. Carr et al. 1993) includes the R39–R62 lines which approach and retreat from the bandhead located at a rest wavelength of 2.29353 $\mu$m (Fig. 1). As described by Carr et al. (1993), the bandhead emission from WL 16 displays the characteristic shape of emission from a rotating disk: a blue wing, a shoulder, and an intensity peak redward of the rest wavelength of the bandhead. That this is the general shape expected for bandhead emission from a rotating disk can be demonstrated by imagining the convolution of the profile of an isolated line from a rotating disk with the distribution of the lines near the $v$=2-0 bandhead (see also Carr 1995).

This convolution procedure is shown schematically in Figure 2 where the double-horned rotational broadening function (heavy line) is the profile of an isolated line from an inclined Keplerian disk that has a monotonically decreasing intensity distribution between an inner and outer radius. The rest spectrum of lines near the bandhead (light line) includes line blending due to local line broadening sources and assumes LTE level populations at a typical inner disk temperature. The blue wing is built up as the wing of the isolated line profile moves past the bandhead (Fig. 2a). A shoulder is formed when one of the horns of the isolated line profile reaches the bandhead (Fig. 2b). As the horn moves past the bandhead it encounters other lines longward of the bandhead; the emission consequently remains approximately constant, producing the shoulder. The red peak forms when the other horn of the isolated line profile reaches the bandhead (Fig. 2c). When both horns have moved past the bandhead, the overlap of the isolated line profile with the lines redward of the bandhead produces the slow descent redward of the red peak (Fig. 2d). In the observed bandhead spectrum, the velocity separation between the blue shoulder and the red peak is approximately twice the projected rotational velocity at the outermost radius of the CO emission, and the maximum observed velocity in the blue wing is the projected velocity of CO at the innermost radius of the emission.

The convolution described above is essentially the procedure used by Carr et al. (1993) to estimate the stellar mass and radius of WL 16. While suitable for this purpose, this procedure is approximate in that the relative strengths of the lines near the bandhead are assumed to be independent of disk radius. The dependence of the emergent spectrum on radial variations of temperature and column density enters only in the rotational broadening function which is effectively the isolated profile of an "average" line at the bandhead. The present modeling procedure removes both of these approximations and allows us to place physical constraints on the properties of the inner disk region.

### 3.2. Application to WL 16

WL 16 is a low luminosity YSO in the $\rho$ Oph cloud ($d$=160 pc). Although originally classified as an embedded (Class I) object (Wilking, Lada, & Young 1989), the measurement of strong 3.3 $\mu$m aromatic hydrocarbon emission (Tokunaga et al. 1991) and a preliminary analysis of our observations of the $v$=2-0 bandhead emission (Carr et al. 1993) imply an earlier spectral type and higher mass for WL 16 than is usually ascribed to Class I objects (e.g., Adams et al. 1987). The bolometric luminosity of WL 16 has been estimated as $18 - 22\,L_\odot$ (Wilking et al. 1989; Cabrit & André 1991). Carr et al. (1993) estimate the extinction to WL 16 to be $A_K = 2$ based on the observed ($J$-$H$) color of WL 16 and assuming that its intrinsic color is similar to that of very active pre-main-sequence stellar systems.

The analysis by Carr et al. (1993) of the $v$=2-0 bandhead emission revealed that parameters of $M_* \simeq 2.5\,M_\odot$, $R_* \leq R_{\rm in} \simeq 5\,R_\odot$, and $i \simeq 60°$ are required to fit both the observed velocity extent and flux of the bandhead emission. As discussed by these authors, an inclination of $i = 60°$ represents a compromise between the projected emitting area required to fit the bandhead flux and the projected rotational velocities required to fit the velocity extent of the emission. Several general considerations constrain the stellar mass to be $\sim 2.5\,M_\odot$ given the assumed extinction to the source. On the one hand, the stellar mass cannot much exceed $\sim 2.5\,M_\odot$ for any pre-main-sequence age given the bolometric luminosity of the system. On the other hand, a smaller stellar mass would require that the emission arise closer to the star to maintain the velocity extent of the bandhead; an origin at smaller radii will tend to reduce the strength of the emission since, for a given bandhead shape, the emission scales as $R_{\rm in}{}^2$. The impact of the reduced emitting area on the bandhead flux can be offset only to a limited extent by adopting a higher temperature for the emitting gas since collisional dissociation of CO becomes important above temperatures of $\sim 5000$K.

To illustrate the additional information that can be extracted from the WL 16 $v$=2-0 data with the present spectral synthesis method, we reanalyze the data using the two-layer LTE model described above. We assume that nearly all of the system luminosity arises from the star and adopt $d = 160$ pc, $i = 60°$, $M_* = 2.5\,M_\odot$, $A_K = 2$, and $L_* = 22\,L_\odot$; the remaining parameters which describe the properties of the disk—$T_{c0}$, $p_c$, $T_{\ell 0}$, $p_\ell$, $\sigma_{\ell 0}$, $q_\ell$, and $\phi_\nu$—are varied to fit the observed bandhead spectrum. For the continuum layer, we adopt a radial temperature dependence in which $p_c = 3/4$, as expected for an active viscous or passive reprocessing disk. The adopted disk continuum temperature corresponds to a disk luminosity of

$$L_D = 2 \int_{R_{\rm in}}^{\infty} \sigma T_c^4 2\pi r\, dr = 4\pi r_0^2 (r_0/R_{\rm in})\sigma T_{c0}^4 \qquad (2)$$

and implies a disk mass accretion rate which can be obtained from the relation

$$\frac{GM_* \dot{M}_D}{2R_{\rm in}} = L_D - \mathcal{F}(1-\mathcal{A})L_* \qquad (3)$$

where the second term accounts for passive reprocessing of incident stellar radiation (Adams & Shu 1986). The quantity $\mathcal{F}$ is the fraction of the stellar luminosity $L_*$ that is intercepted by the disk ($\mathcal{F} = 1/4$ if $R_{\rm in} = R_*$) and $\mathcal{A}$ is the effective scattering albedo. Values for the other disk parameters $T_{c0}$, $T_{\ell 0}$, and $\sigma_{\ell 0}$ are chosen to reproduce both the strength of the 2.3$\mu$m continuum and the contrast of the bandhead above the continuum. More detailed spectral features such as the shape of the blue wing, the sharpness of the shoulder, the location of the red peak, and the spectral shape redward of the peak constrain $p_\ell$ and $q_\ell$.

With these assumptions, we find that we can fit the strength of the 2.3$\mu$m continuum if the $M_* = 2.5\,M_\odot$, $L_* = 22\,L_\odot$ star lies near the stellar birthline ($R_* = 5.0\,R_\odot$, Palla & Stahler 1993) and the continuum layer has a temperature of $T_{c0} = 3000$K at $r_0 = 5R_\odot$. In order to fit the velocity extent and strength of the blue wing, $T_\ell$ must be $\lesssim 5000$K at $5\,R_\odot$. These high temperatures correspond to large line source functions without appreciable dissociation of CO, and thereby maximize the bandhead flux from these radii. The location of the shoulder and red peak indicates that most of emission arises from within 30 $R_\odot$; the sharpness of these features indicates that the emission must decrease relatively sharply beyond this radius. If the rapid decrease in the emission beyond $30R_\odot$ is due, in part, to the reduced sensitivity of the $v$=2-0 bandhead lines to gas below 1500K (see below), the constraints on the line layer temperature are roughly $T_\ell \simeq 1500$ K at $\sim 30\,R_\odot$ and $T_\ell \simeq 5000$ K at $\sim 5\,R_\odot$ which corresponds to a temperature profile with $T_{\ell 0} \simeq 5000$ and $p_\ell \simeq 3/4$. Flatter temperature profiles that roughly preserve the locations of the shoulder and red peak and velocity extent of the the blue wing have lower temperatures within $\sim 30R_\odot$ and require somewhat steeper column density distributions to fit the sharpness of the shoulder and red peak.

Given these restrictive requirements, it is very difficult to fit the strength and shape of the bandhead with thermal Doppler profiles. Within the context of the two-layer LTE model, we can obtain a good fit only within a very restricted range of parameters about $\sigma_{\ell 0} = 5000\,{\rm g\,cm^{-2}}$, $q_\ell = 4.5$, $T_{\ell 0} = 6000$K, $p_\ell = 3/4$, and $R_{\rm in} = R_*$ (Model L12; see Fig. 3a and Table 1). Since the wings of gaussian profiles are weak, it is difficult to emit much flux between the line cores without these large column densities and the higher line layer temperature. These column densities are, in fact, so large as to violate our assumption of a negligible 2.3$\mu$m continuum optical depth $\tau$ in the line layer: the opacity due to H$^-$ alone corresponds to $\tau = 700$ at $1.1R_*$ and $\tau > 1$ within $1.8R_*$.

Including this continuum optical depth in the model would significantly worsen the fit. In Figure 3a, we are able to fit the strength and extent of the blue wing and red peak because the high column density close to the star yields a sufficient abundance of CO to keep the optical depth of the overtone lines appreciable even at $T_\ell = 6000$K. If we were to include the line layer continuum optical depth in the model, the high optical depth close to the star would significantly reduce the contrast of the line emission from this region and result in an excessively large 2.3$\mu$m continuum as well as both an inadequate velocity extent and overall strength for the bandhead emission (cf. Fig. 3d). In general, somewhat smaller line layer column densities can produce the same bandhead flux if the line layer represents the entire disk column density within a few stellar radii, i.e., if the continuum layer is absent. In the present case, however, even the absence of a continuum layer cannot sufficiently reduce the required column density to values corresponding to reasonable H$^-$ opacities within $\sim 1.6R_*$. We conclude that, given the adopted extinction to the source, thermally broadened lines cannot account for the CO emission from WL16.

Alternatively, line layers with column densities of $\sim 100\,\mathrm{g\,cm^{-2}}$ at $10R_\odot$ and underlying optically thick continuum layers can produce good fits to the strength of both the bandhead and the 2.3$\mu$m continuum if the intrinsic line profiles are broader than thermal. With Lorentzian lines, these fits require systematically lower column densities than in the case of thermal Doppler lines because more flux can be emitted in the stronger wings of Lorentzian lines. Figure 3b (Model L14) shows the fit obtained with LTE populations, Lorentzian line profiles with a constant line width (half-width at half maximum) of $\gamma = 0.5\,\mathrm{km\,s^{-1}}$, $\sigma_{\ell 0} = 370\,\mathrm{g\,cm^{-2}}$, $q_\ell = 3.8$, $p_\ell = 0.78$, and all other parameters the same as in Model L12 (cf. Table 1). At the lower line layer column densities of this model, the CO dissociation radius lies at a somewhat larger distance than in Model L12: in Model L14 the abundance and optical depth in the overtone lines become substantial beyond $1.2R_*$. Beyond this distance, the continuum optical depth due to H$^-$ is negligible, consistent with our assumptions. We are able to fit the extent of the blue wing despite the larger dissociation radius because of the stronger wing of the Lorentzian line. With these parameters, the shadowing of the star by the disk reduces the stellar contribution to the 2.3$\mu$m continuum to $\sim 0.29$ Jy. The star, in turn, shadows the part of the disk that contributes line emission at low projected velocities, enhancing the sharpness of the shoulder. The (partially shadowed) disk continuum layer contributes an additional $\sim 0.12$ Jy to the 2.3$\mu$m continuum and $\sim 2\,L_\odot$ to the system luminosity (eq. 2).

Because the lines in this model are much less optically thick than in the previous case, there is a larger region of parameter space within which we can find acceptable model fits. For example, a similar fit is possible with a column density distribution as flat as $q_\ell = 2$ (with $\sigma_{\ell 0} = 90\,\mathrm{g\,cm^{-2}}$, $T_{\ell 0} = 5500$K, and $p_\ell = 3/4$). Temperature distributions as flat as $p_\ell = 0.5$ can also fit the data if the column density distribution is steeper (e.g., $\sigma_{\ell 0} = 400\,\mathrm{g\,cm^{-2}}$, $q_\ell = 4.5$, and $T_{\ell 0} = 4700$K). Larger Lorentzian linewidths further decrease the column density required to produce the strength of the bandhead and tend to flatten the slope of the spectrum redward of the peak. Using somewhat steeper temperature gradients to compensate for this flattening will produce a good fit. For example, for $\gamma = 2.5\,\mathrm{km\,s^{-1}}$, adopting $T_{\ell 0} = 5900$K, $p_\ell = 0.78$, $\sigma_{\ell 0} = 75\,\mathrm{g\,cm^{-2}}$, and $q_\ell = 3$ will also fit the data.

WL 16 is unusual among low luminosity YSOs in showing strong aromatic hydrocarbon emission features (e.g., Hanner et al. 1992), a property that has been interpreted as evidence for an early stellar spectral type. High resolution $H$-band spectroscopy of WL 16 is also consistent with an early spectral type (see §3.4). The system luminosity of $22L_\odot$ and deduced stellar mass restrict the evolutionary state of such a star. In order to investigate the possibility of an earlier stellar spectral type for WL 16, we consider whether WL 16 is an A-star with mass $M_* = 2.5M_\odot$, radius $R_* = 2.08R_\odot$, and luminosity $L_* = 22L_\odot$. Figure 3c (Model L15) shows the synthesized spectrum obtained with these stellar parameters, Lorentzian line profiles, and the following set of disk parameters: $r_0 = 5R_\odot$, $T_{c0} = 3000$K, $T_{\ell 0} = 5810$K, $p_\ell = 0.78$, $\sigma_{\ell 0} = 310\,\mathrm{g\,cm^{-2}}$, $q_\ell = 3.7$, and $\gamma = 0.5\,\mathrm{km\,s^{-1}}$ (cf. Table 1). The star contributes $\sim 0.1$ Jy of the continuum; the disk extends up to the surface of the star and contributes $\sim 0.2$ Jy to the 2.3$\mu$m continuum and $\sim 4\,L_\odot$ to the system luminosity. The (inner) dissociation radius is effectively $\sim 5\,R_\odot$, as in Model L14. It would be difficult to distinguish between this dissociation radius and a disk truncation radius $R_{\mathrm{in}} \lesssim 5\,R_\odot$ because the additional (unshadowed) stellar contribution to the continuum in the latter case is comparable to the contribution to the continuum from the truncated disk region. Although the fit to the bandhead emission is comparable to that obtained with Model L14, the continuum is reduced by a third due to the earlier stellar spectral type.[2]

### 3.3. Discussion of WL 16 Results

The detailed modeling procedure presented above confirms the result obtained by Carr et al. (1993) that the parameters $M_* = 2.5\,M_\odot$, $R_* \leq R_{\rm in} = 5\,R_\odot$, and $i = 60°$ provide a good fit to both the observed velocity extent and flux of the bandhead emission. We are able to explain the inner disk radius of $5\,R_\odot$ as either a CO dissociation radius (Model L15) or a physical truncation radius. The disk may either extend up to the surface of a $5\,R_\odot$ star on the stellar birthline (Model L14) or it may be truncated due to the presence of a stellar magnetosphere that extends to a similar distance (a variation of Model L15).

By fitting the detailed shape of the bandhead emission, e.g., the location and sharpness of the shoulder, we find that that the $v$=2-0 emission must be truncated relatively rapidly beyond $30\,R_\odot$. This effect can arise for a variety of reasons and probably does not indicate a physical outer edge to the disk. For example, at temperatures below 1500K, the lines that make up the $v$=2-0 bandhead are not well excited even if the level populations are in LTE. In the models shown, a steep column density gradient is used to emphasize this outer radius to the emission. Flatter column density distributions are possible if dust is present in the upper disk atmosphere (i.e., the line layer) below 1500K. Because the CO lines and the 2.3$\mu$m continuum would then form in the same layer this would eliminate the contrast of the CO emission above the continuum beyond this radius, creating an effective outer radius to the emission. The role of non-LTE effects in creating effective outer radii to the overtone emission are discussed in the next section.

We find that suprathermal line widths are needed to fit the strength and shape of the bandhead emission. The bandhead emission is well fit with Lorentzian lines characterized by $\gamma = 0.5 - 2.5\,{\rm km\,s^{-1}}$. In §6 we offer a physical interpretation for these enhanced line widths. We also find that the required line layer column densities ($\sim 100\,{\rm g\,cm^{-2}}$ at $10\,R_\odot$) are much lower than the values ($\sim 10^5\,{\rm g\,cm^{-2}}$) expected by extrapolating models of the "minimum solar nebula" to these radii. We interpret this result as indicating that the line layer represents a temperature inversion region in the upper atmosphere of a disk that has a much larger total column density, and note that this result also indicates the sensitivity of the CO overtone lines to relatively small amounts of circumstellar matter. Similar column densities may be present in the continuum "gaps" or "inner holes" of YSO disks (cf. Strom et al. 1989; Hillenbrand et al. 1992; Lada & Adams 1992; Marsh & Mahoney 1992; Mathieu 1994).

The disk continuum temperature used in all fits corresponds to a disk mass accretion rate that is consistent with zero but is in all cases $\lesssim 2 \times 10^{-7}\,M_\odot\,{\rm yr}^{-1}$. The $2L_\odot$ radiated by the disk in Model L14 can be entirely attributed to reprocessed starlight if the effective albedo is 0.68 (eq. 3 with $\mathcal{F} = 1/4$). Similarly, the $4L_\odot$ disk luminosity in Model L15 could be due to reprocessed starlight alone if the effective disk albedo is 0.2. Alternatively, the disk luminosity in both models can be attributed to energy release by disk accretion at $\sim 2 \times 10^{-7}\, M_\odot\, \mathrm{yr}^{-1}$ if the effective albedo is unity. This accretion rate is low compared to typical values for both Class I objects (e.g., Adams et al. 1987) and Group II Herbig Ae stars (Hillenbrand et al. 1992), consistent with the lack of evidence for either an outflow from WL 16 or other phenomena commonly associated with active Class I sources such as SVS 13 and L1551 IRS5.

---

[2] Note that some fraction of the 2.3$\mu$m continuum has been added "by hand" to produce the synthesized spectra shown in Figures 3a,b,c (Models L12, L14, L15). The discrepancy is minor in the case of Models L12 and L14 but accounts for $\sim 1/3$ of the continuum in Model L15.

The synthetic spectra shown in this section indicate the degree to which modeling the $v$=2-0 bandhead spectral region assuming LTE level populations allows us to constrain the physical properties of the disk. We find that there is some trade off between the line layer temperature and column density profiles required to obtain a good fit. This non-uniqueness is due to both the assumption of LTE and the choice of the $v$=2-0 bandhead as the region to model. Since the assumption of LTE fixes the source function, this leaves the optical depth as the only variable in the radiative transfer that is affected by a given temperature and column density. Since the lines in the $v$=2-0 bandhead region are excited at similar temperatures and moreover probe a restricted range in optical depth at a given temperature and column density, the trade off between temperature and column density in the optical depth will work almost equally well for all the lines in the spectral region, producing a limited effect on the shape of the bandhead. In comparison, the consideration of non-LTE effects and spectral fitting of multiple bandhead regions provide powerful additional diagnostics of the detailed physical properties of the inner disk. Following a brief discussion of the evolutionary status of WL 16 (§3.4), we investigate these additional diagnostics. In § 4, we construct a non-LTE model and in §5 we apply the model to $v$=2-0 and $v$=5-3 observations of another YSO, 1548C27, in order to demonstrate the constraints that these additional considerations allow us to place on the physical properties of inner YSO disks.

### 3.4. WL 16: On the Stellar Birthline or Main-sequence Star?

In fitting the $v$=2-0 bandhead emission, we find that we can model the stellar component of WL 16 as either a G-star on the birthline or an A-star near the main-sequence. The presence of aromatic hydrocarbon features in the spectrum of WL 16 and the fit to the 2.3$\mu$m continuum provide other potential discriminants between these two possibilities. Of the various aromatic hydrocarbon features observed in the spectrum of WL 16, the 3.3$\mu$m feature is the strongest indicator of an early spectral type because it is excited by photons shortward of 3000Å (Schutte et al. 1993). We can compare the efficiency required to excite the observed 3.3$\mu$m feature with UV radiation from either stellar spectral type with the efficiencies measured for other astronomical sources. For the models considered above, the fraction of the stellar (blackbody) luminosity that is emitted between 912Å and 3000Å, is 1% for $T_* = 5600$K (Model L14) and 6% for $T_* = 8700$K (Model L15). For $d = 160$ pc, the strength of the 3.3$\mu$m feature in WL 16 is $\sim 2 \times 10^{-3}\, L_\odot$ (cf. Tokunaga et al. 1991) where we have assumed an extinction correction at 3.3$\mu$m appropriate for $A_K = 2$. This luminosity is 0.9% of the available UV radiation for a 22$L_\odot$ star with $T_* = 5600$K and 0.14% for $T_* = 8700$K. Although the latter value is more typical of those measured for Herbig AeBe stars, the former is not unreasonable given the large range in the measured values (Brooke et al. 1993). Thus, the strength of the 3.3$\mu$m feature does not strongly imply a spectral type earlier than G.

While the continuum level is fit fairly well with a G-star on the birthline and an optically thick inner disk, if WL 16 is instead an A-star near the main-sequence, we require an additional source of 2.3$\mu$m continuum. The very small grains (VSGs) that are the larger counterparts of aromatic hydrocarbons are likely to accompany these macro-molecules in the near environment of WL 16 and may contribute to this continuum. Natta et al. (1993) have shown that VSGs surrounding Herbig AeBe stars can reprocess stellar radiation into the near- and mid-IR ($\sim 2-20\mu$m), significantly enhancing the strength of the continuum at $2.3\mu$m. Compared to the aromatic hydrocarbons responsible for the $3.3\mu$m feature, VSGs are excited by a much broader spectrum of photons. Thus they may significantly enhance the $2.3\mu$m continuum of WL 16 over a range of stellar spectral types and may make up the deficit of $2.3\mu$m continuum noted above for Model L15.

In an effort to better determine the spectral type of WL 16, we obtained a high-resolution CSHELL spectrum centered on the $1.711\mu$m Mg line. For stars of spectral type G and later, this line is among the strongest stellar absorption features in the $H$ band. While WL 16 is substantially fainter at this wavelength ($H = 10.5$) than in the $K$ band ($K = 7.8$) the detectability of stellar lines should be better since the infrared excess in pre-main-sequence stars generally decreases with decreasing wavelength. The total integration time was 70 minutes through a $2''$ slit, giving a spectral resolution of $\sim 23\,\mathrm{km\,s^{-1}}$ and a signal-to-noise ratio of 70. No absorption features were present. The limiting factor on a line detection is the residual fringing in the spectrum which has a peak-to-peak amplitude of 4% and a period of about $360\,\mathrm{km\,s^{-1}}$.

The non-detection of the Mg line could be due to any of the following reasons: the infrared continuum excess significantly dilutes the stellar spectrum; the stellar rotational broadening of the line is large; the star is hot enough that the line is weak or absent; or some combination of these. In order to investigate the constraints that can be placed on the spectral type of WL 16 from the $H$-band data, we calculated synthetic stellar spectra for the 1.711 $\mu$m Mg line. We used an updated version of the spectrum synthesis program moog (Sneden 1973) and the Kurucz grid of model atmospheres (Kurucz 1993). The $gf$-value and damping constants for the Mg line were determined by fitting an observed solar intensity spectrum (Livingston & Wallace 1991).

In the G star model for WL 16 described in §3.2 (Model L14) $\sim 70$% of the $2.3\mu$m continuum comes from the star; the star should contribute an even larger fraction of the continuum at $1.7\mu$m. For the assumed stellar parameters ($T_\mathrm{eff} = 5500$K and $\log g = 3.5$), the Mg line is predicted to be 35% deep at our spectral resolution in the absence of rotational broadening and other sources of continuum emission. In order for rotational broadening alone to account for the non-detection of the Mg line, we require $v \sin i \geq 200\,\mathrm{km\,s^{-1}}$ which is a significant fraction of the breakup velocity for such a star ($310\,\mathrm{km\,s^{-1}}$). A larger $1.7\mu$m continuum excess reduces the required rotational velocity. For example, with a more moderate $v \sin i$ of $50\,\mathrm{km\,s^{-1}}$ we we require an excess flux of at least 3 times the stellar flux. We conclude that a late-type star cannot provide the major contribution to the near-infrared continuum unless the star is an extremely rapid rotator.

The Mg line strength decreases with increasing effective temperature. For the temperature of our A star model (Model L15, $T_\mathrm{eff} = 8500$K) the observed depth is predicted to be 17%. In this model, the A star contributes only 23% of the $2.3\mu$m continuum flux. Assuming that the star contributes $\sim 35$% at $1.7\mu$m , a rather low $v \sin i$ of $25\,\mathrm{km\,s^{-1}}$ would be sufficient to hide the line in our data. Thus an A star, with the majority of the continuum coming from excess emission, would be consistent with the non-detection of the Mg line.

## 4. Non-LTE Model and Multi-wavelength Fits
*4.1. Level Population Calculation*

We calculate the population of the rovibrational levels using a ten-level CO molecule to represent the vibrational levels and assuming that the rotational levels within each vibrational level are thermally populated. This approximation is based on the relatively large rate coefficients for the collisional de-excitation of pure rotational transitions (Green & Thaddeus 1976, Flower & Launay 1985, Schinke et al. 1985). In the temperature range of interest, the collisional de-excitation rate for both H and $H_2$, $k(v, J \to v, J - \Delta J) \sim 5 \times 10^{-10}$ cm$^3$ s$^{-1}$ for $\Delta J = 1, 2, \ldots$, is much larger than the corresponding rate coefficients for collisional changes in the vibrational (as well as rotational) quantum numbers. This approximation is valid even in the important case in which collisions with atomic hydrogen dominate the collisional rates of the rovibrational transitions, where $k(v, J \to v - 1, J - \Delta J) \sim 10^{-11}$ cm$^3$ s$^{-1}$ (cf. Appendix B).[3] The assumption of rotational equilibrium greatly simplifies the determination of the level populations and allows us to include in a simple way the effects of radiative trapping on the level populations. In this case, the vibrational level populations can be determined by the solution of a coupled set of effective vibrational level population equations which we derive in the following way.

The large oscillator strengths and collision cross-sections of the $\Delta v = 1$ transitions suggest that the population of a given vibrational level is primarily determined by radiative and collisional interactions with adjacent vibrational levels. Therefore, we begin by assuming that each vibrational level $v$ is directly coupled only to the vibrational level above it, $v' = v + 1$. As discussed by Scoville et al. (1980), in this approximation successively higher vibrationally excited states are populated by climbing the vibrational ladder "one rung at a time". In steady state, the number of transitions into level $v$ from level $v'$ is balanced by the number of transitions into level $v'$ from level $v$ :

$$\sum_{u,\ell} n_u (A_{u\ell} + B_{u\ell} \overline{J}_\nu + C_{u\ell}) = \sum_{u,\ell} n_\ell (B_{\ell u} \overline{J}_\nu + C_{\ell u}), \qquad (4)$$

where we have adopted the notation $\ell = (v, J)$ and $u = (v', J')$ and the sum is over all transitions permitted between the upper and lower vibrational levels (e.g., the R and P branches). In this equation, $n_u = n(v', J')$ and $n_\ell = n(v, J)$ are the number densities of the upper and lower levels; $A_{u\ell}$, $B_{u\ell}$, and $B_{\ell u}$ are the Einstein A's and B's of the transitions connecting the upper and lower levels; $C_{\ell u}$ and $C_{u\ell}$ are the upward and downward collisional rates for these transitions assuming a sum over all collision partners (H, $H_2$, e, *etc.*). We have also denoted the mean intensity of the radiation field weighted by the line profile function $\phi_\nu$ as $\overline{J}_\nu \equiv \int J_\nu \phi_\nu d\nu$.

---

[3] There is, in principle, an important difference between the rotational equilibrium of ground and excited vibrational levels stemming from the Einstein A-values for the two cases. Since the A-value is proportional to the cube of the transition frequency, the critical density for an excited vibrational level can be very large, typically $n_{\mathrm{crit}}(v > 0, J) \sim 10^{10}$ cm$^{-3}$. Nevertheless, the large difference between vibrational and rotational collisional rates means that the total vibrational populations are more readily disequilibrated, and we choose to focus on this aspect of a non-LTE calculation.

Consistent with our assumption of vertical homogeneity (e.g., of the level populations) in the line layer, we calculate $\overline{J}_\nu$ at an "average" height in the slab, located where the vertical optical depth is half of the total. In addition, we ignore the effect of overlapping lines in the determination of the mean intensity. While the issue of overlapping lines is important for the calculation of the intensity *emergent* from the slab, we show below that the line overlap does not significantly affect the level populations. Thus, in the calculation of the level populations we treat each line as isolated and assume that the mean intensity which excites the line is a function of the optical depth for that line alone (i.e., $\overline{J}_\nu = \overline{J}_\nu^{u\ell}$).

With these assumptions, $\overline{J}_\nu$ for a given line is (cf. eq. 1)[4]

$$\overline{J}_\nu \equiv \frac{1}{4\pi}\int \phi_\nu d\nu \int I_\nu d\Omega = B_\nu(T_c)\frac{1}{2}\int_{-\infty}^{\infty}\phi_x dx \int_0^1 e^{-\overline{\tau}_\nu \phi_x/2\mu}d\mu$$
$$+ S_\nu\left(1 - \frac{1}{2}\int_{-\infty}^{\infty}\phi_x dx \int_{-1}^1 e^{-\overline{\tau}_\nu \phi_x/2|\mu|}d\mu\right)$$

where all physical quantities such as $B_\nu(T_c)$, $S_\nu$, and $\overline{\tau}_\nu$ are assumed to be independent of $\mu$ because the disk is geometrically thin. The integration limits reflect the difference between the background radiation field which originates from the $2\pi$ steradians subtended by the continuum layer and the diffuse field which originates from all $4\pi$ steradians. The optical depth

$$\overline{\tau}_\nu = \frac{1}{\Delta\nu}\int \tau_\nu d\nu \qquad (5)$$

is an average over the line profile where $\Delta\nu$ is the characteristic width of the line profile centered on the line frequency $\nu_0$.

We introduce a dimensionless line profile function $\phi_x \equiv \phi_\nu \Delta\nu$ such that $\tau_\nu = \overline{\tau}_\nu \phi_x$ and $x \equiv (\nu - \nu_0)/\Delta\nu$. Following Mihalas' treatment of the escape probability in a static plane parallel atmosphere with vertically homogeneous properties (Mihalas 1978), we interpret quantities of the form

$$\beta(\tau) = \int_{-\infty}^{\infty}\phi_x\, dx \int_0^1 e^{-\tau\phi_x/\mu}d\mu = \int_{-\infty}^{\infty}E_2(\tau\phi_x)\phi_x\, dx \qquad (6)$$

as escape probabilities. The line profile function in the outer integral expresses the probability that at a vertical optical depth $\tau$ a photon is emitted $x = (\nu - \nu_0)/\Delta\nu$ line widths from line center. Such a photon has a probablility $E_2(\tau\phi_x)$ of escaping the layer in any direction $\mu$. Our expression for $\beta(\tau)$ differs from Mihalas' expression for $P_e(\tau)$ by a factor of 2 because we have assumed that the photon can escape out of both the top and bottom of the atmosphere and that the optical depth in either direction is $\tau = \overline{\tau}_\nu/2$ on average. While the photon cannot really "escape" by propagating into the continuum layer below, we imagine that photons reaching the continuum layer are destroyed and reemitted as continuum radiation. With this identification, the mean intensity weighted by the line profile is

$$\overline{J}_\nu = B_\nu(T_c)\beta(\overline{\tau}_\nu/2)/2 + S_\nu\left[1 - \beta(\overline{\tau}_\nu/2)\right]. \qquad (7)$$

---

[4] In the interest of notational economy, we have dropped the superscripts $u\ell$ on quantities such as $\overline{J}_\nu$, $I_\nu$, $S_\nu$, $\tau_\nu$ and $\beta$.

Note that the limits of integration in dimensionless frequency $x$ in equation (6) assume an isolated line. For the CO fundamental transitions, the P-transitions are separated by 4-8 cm$^{-1}$ and the R-transitions are more closely spaced (0-6 cm$^{-1}$) since the transitions go through a bandhead. If we conservatively estimate the average spacing between neighboring lines to be $\Delta\tilde{\nu} = 1\,\text{cm}^{-1}$, truncating the integral at $x_t = \Delta\tilde{\nu}/\delta\tilde{\nu}$ ($\simeq 50$ for a linewidth $\delta\tilde{\nu}$ of $2.5\,\text{km s}^{-1}$) reduces the escape probability by less than 50% relative to the untruncated case. So overlapping lines do not have a dominant effect on the relevant optical depths and escape probabilities.

Using the definition of the line source function to rewrite equation (7) for a given line as

$$\overline{J}_\nu^{u\ell} = \frac{A_{u\ell}}{B_{u\ell}} \left[ \frac{(1-\beta_{u\ell})}{n_\ell g_u/n_u g_\ell - 1} + \frac{\beta_{u\ell}/2}{e^{h\nu/kT_c} - 1} \right]$$

where $\beta_{u\ell} = \beta(\overline{\tau}_\nu^{u\ell}/2)$ and $n_\ell$, $n_u$, $g_\ell$, and $g_u$ are the number densities and statistical weights of CO in the lower and upper levels, equation (4) becomes

$$\sum_{u\ell} (n_\ell C_{\ell u} - n_u C_{u\ell}) = \sum_{u\ell} n_u A_{u\ell} \beta_{u\ell} \left[ 1 - \frac{(n_\ell g_u/n_u g_\ell - 1)/2}{e^{h\nu/kT_c} - 1} \right]. \tag{8}$$

If the rotational levels are thermally populated at the gas kinetic temperature $T_\ell$

$$\frac{n_{vJ}}{n_v} = \frac{(2J+1)}{Z} e^{-J(J+1)B/T_\ell} \equiv f_J, \tag{9}$$

where $n_{vJ} = n(v,J)$, $n_v = \sum_J n_{vJ}$, $B$ is the rotational constant, and $Z \simeq T_\ell/B$ is the partition function for the rotational levels, the radiative terms in equation (8) can be further approximated as

$$\left[ 1 - \frac{(n_v/n_{v'} - 1)/2}{e^{h\nu/kT_c} - 1} \right] \sum_{u\ell} n_u A_{u\ell} \beta_{u\ell}.$$

With this approximation, the assumption of thermally populated rotational levels, and a little algebra, equation (8) becomes

$$n_{v'} \left[ C^*_{v'v} + \left( 1 + \frac{1/2}{e^{h\nu/kT_c} - 1} \right) A^*_{v'v} \beta^*_{v'v} \right] = n_v \left[ C^*_{vv'} + \frac{1/2}{e^{h\nu/kT_c} - 1} A^*_{v'v} \beta^*_{v'v} \right] \tag{10}$$

where

$$A^*_{v'v} \equiv \sum_{u\ell} f_{J'} A_{u\ell}, \qquad C^*_{v'v} \equiv \sum_{u\ell} f_{J'} C_{u\ell},$$
$$\beta^*_{v'v} \equiv \frac{1}{2} \sum_{u\ell} f_{J'} \beta_{u\ell}, \qquad C^*_{vv'} \equiv \sum_{u\ell} f_J C_{u\ell}. \tag{11}$$

The terms $A^*_{v'v}$, $\beta^*_{v'v}$, $C^*_{v'v}$, and $C^*_{vv'}$, can be regarded as effective Einstein A's, escape probabilities, and collisional terms that relate the vibrational levels. Note that we have made the additional approximation of separately averaging the A-values and escape probabilities since the A-values usually do not vary substantially over the range of $J$ in which the escape probabilities are substantial. The factor of 1/2 in the definition of $\beta^*_{v'v}$ arises from the consideration of the R and P transitions out of the upper levels.

By detailed balance the upward and downward collision rates satisfy $C_{J'J} = (n_J/n_{J'})_{\text{LTE}}\, C_{JJ'}$. The assumption of thermal rotational level populations now implies that

$$C^*_{v'v} = e^{(v'-v)\,\theta/T}\, C^*_{vv'}$$

where $\theta$ is the vibrational constant. When collisions dominate, equations (10) and (12) yield the LTE result: $n_{v'}/n_v = \exp[-(v'-v)\theta/T_\ell]$. When radiative transitions dominate, equation (10) implies $n_{v'}/n_v = 1/[2\exp(h\nu/kT_c) - 1]$, which differs from the Boltzmann relation for excitation temperature $T_c$ because the underlying radiation field is, as noted earlier, only half a blackbody.

The system of coupled, statistical-equilibrium equations for the vibrational level populations $n_v$ is linked to the radiative transfer part of the problem through the determination of the escape probabilities $\beta_{u\ell} \equiv \beta(\overline{\tau}^{u\ell}_\nu/2)$ which requires the self-consistent determination of the vibrational level populations that enter in the optical depths $\overline{\tau}^{u\ell}_\nu$. This suggests that the system of equations can be solved iteratively given the initial assumption of LTE level populations. The rates for vibrational excitation due to collisions with H and $H_2$ are based on the rate coefficients discussed in Appendix B. The total collision rates, $C^*_{v'v}$ and $C^*_{vv'}$, are obtained by assuming chemical equilibrium abundances of H and $H_2$. Although Einstein A-values for individual lines within a vibrational band can vary by factors of several, the average A-value between two vibrational levels $A^*_{v'v}$ varies weakly with temperature (see also Appendix A). Therefore, we adopt a temperature-independent rate of $A^*_{v'v} = A^*_{v'v}(T = 2000K)$. The set of equations (10) can easily be generalized to couple more than two neighboring vibrational levels together. It is usually not necessary to couple more than three levels together to solve the set of equations (10) to a fractional accuracy of $< 10^{-7}$.

### 4.2. Spectral Synthesis of the v=5-3 and v=2-0 Bandhead Regions

The $v$=5-3 bandhead region of the spectrum provides valuable complementary information to that contained in the $v$=2-0 bandhead region. For example, the onset of non-LTE level populations is difficult to detect with measurements of the $v$=2-0 bandhead alone. This is evident from a comparison of the fits to the WL16 $v$=2-0 bandhead region shown in Figures 3b-c and 3e-f. The latter set of models fits, obtained with the non-LTE model described in the previous section, show that similar fits to the data can be obtained with the nearly the same parameters as in the LTE case (cf. Table 1). This is because the $v$=2-0 bandhead region consists of lines that are primarily sensitive to gas at temperatures larger than those at which departures from LTE become important. This is shown graphically in Figure 4 in terms of the relative optical depth of a line at the $v$=2-0 bandhead in LTE as a function of temperature. Because lines in the $v$=2-0 bandhead arise from relatively high-lying rotational levels, they are not well excited below 1500 K even in LTE. Thus, while the lines in the $v$=2-0 bandhead provide useful information about the kinematics of the gas in the inner disk, they probe only a limited range in temperature.

In comparison, isolated lines from several other CO overtone transitions also fall in the $v$=5-3 bandhead region (Fig. 5); these lines and the 5-3 bandhead together probe a much wider range of excitation temperatures (500K<T<5000K) than those contributing to the $v$=2-0 bandhead. In Figure 4, the relative LTE optical depths $f_{\ell u}(2J_\ell + 1)\exp(-E_\ell/kT)$ of representative lines from different overtone transitions in the $v$=5-3 bandhead region are compared with that of a line at the $v$=2-0 bandhead. The strongest lines in the $v$=5-3 bandhead region are the 2-0 and 4-2 lines and the 5-3 bandhead itself. The 2-0 P transitions are the primary low temperature constraint. Because the 4-2 and 2-0 lines do not strongly overlap, it is possible to establish their relative strengths. The relatively large separation in wavelength between these lines also makes it possible to measure the line profiles of individual lines for sources with projected rotational velocities $\lesssim 100$ km s$^{-1}$. In the next section, we present our high spectral resolution observations and modeling of the $v$=2-0 and $v$=5-3 bandhead emission from such a source, the probable Herbig AeBe star 1548C27.

## 5. $v$=5-3 and $v$=2-0 Bandhead Spectroscopy of 1548C27
### 5.1. What is 1548C27?

Discovered as a cometary nebula (Craine et al. 1981), the YSO 1548C27 has an associated jet which is aligned with the nebular axis (Mundt et al. 1984). Energetic outflow activity is also indicated by the optical spectrum of the central object which is characterized by P Cygni H$\alpha$, H$\beta$ and broad blueshifted NaI D absorption (Mundt et al. 1987). In the near-infrared, 1548C27 has very prominent CO overtone and Br $\gamma$ emission (Carr 1989, 1990). The $U$-band – 100 $\mu$m spectral energy distribution indicates a system luminosity of $L \simeq 128\,(d/\text{kpc})^2\,L_\odot$ (Vilchez et al. 1989) with the spectral shape indicating Herbig AeBe Group II membership (cf. Hillenbrand et al. 1992).

The distance to 1548C27 is controversial. It is located in front of the HII region S86 and in the same region of the sky as NGC 6823, both of which are located at a distance of $\sim$ 2.5 - 3 kpc. Arguments based on the spectral energy distribution shortward of 3 $\mu$m, suggest that 1548C27 is at a distance of $d = 0.91$ kpc while CO ($J$=2-1) and CO ($J$=3-2) data have been used to derive a kinematic distance of 2.4 kpc (Dent & Aspin 1992).

### 5.2. CO Overtone Spectroscopy

Spectra of 1548C27 in the $v$=2-0 and $v$=5-3 bandhead regions were obtained on 1993 May 15 at the NASA Infrared Telescope Facility using the facility cryogenic echelle spectrograph (CSHELL; Tokunaga et al. 1990, Greene et al. 1993) with a NICMOS 3 256×256 HgCdTe array and a 1″ slit width that provided a resolution of 14 km s$^{-1}$. The data were acquired and reduced as described in Carr et al. (1993). The resulting spectra for the $v$=2-0 and $v$=5-3 bandhead regions are shown in Figure 5. As in the case of WL 16, the 2-0 bandhead of 1548C27 also shows the characteristic shape of bandhead emission from a rotating disk. Although the velocity extent of the bandhead emission (100 – 120 km s$^{-1}$) is less than in the case of WL 16, the characteristic blue wing, shoulder and red peak are clearly present.

In the $v$=5-3 bandhead region, the feature that peaks at 2.3837$\mu$m is a complex blend of the 5-3 bandhead, located at 2.3829$\mu$m, and neighboring 4-2 R and 2-0 P lines. Although the kinematics of the emitting gas are difficult to deduce from the shape of this feature, the double-peaked shape of the features at 2.3795 and 2.3810$\mu$m strongly support the rotating disk interpretation for the emission. The feature at 2.3795$\mu$m is essentially a 4-2 R line and the feature at 2.3810$\mu$m is a blend of a 4-2 R line and a 2-0 P line. Based on the velocity extent of the $v$=2-0 bandhead emission, these features overlap only slightly in their respective red and blue wings and are a reasonably good representation of an isolated line profile.

### 5.3. Modeling of the $v$=2-0 and $v$=5-3 spectra

To demonstrate the constraints that these observations place on the inner disk properties of the system, we assume that the distance to 1548C27 is 2.4kpc (cf. Dent & Aspin 1992) and, therefore, the system luminosity is $740 L_\odot$. Studies of the evolution of intermediate mass pre-main sequence stars restrict the stellar mass of such a system to $\lesssim 5 M_\odot$ (Palla & Stahler 1993). For this mass, the velocity extent of the 2-0 bandhead implies that the system must be viewed at relatively low inclination. For a $K$-band extinction of $A_K = 0.62$ mag (Vilchez et al. 1989), and $i = 30°$, we can roughly match the velocity extent of the bandhead emission with $M_* = 4 M_\odot$ and a CO dissociation radius at $\sim 20 R_\odot$. Assuming that the continuum layer has the usual radial temperature dependence $T_c = T_{c0}(r/r_0)^{-3/4}$, we can fit the 2.3 $\mu$m continuum level and the system luminosity with $R_* = r_0 = 5.9 R_\odot$ and $T_{c0} = 9800$K. These parameters correspond to a stellar luminosity of $L_* = 200 L_\odot$, an age of $1.5 \times 10^5$ yr from the birthline (Palla & Stahler 1993), and a disk luminosity of $L_D = 4\pi R_*^2 \sigma T_{c0}^4 = 280 L_\odot$. We assume that a boundary layer/hot spot contributes another 260 $L_\odot$ so that the total system luminosity is $740 L_\odot$.

The adopted stellar parameters place 1548C27 among the Herbig AeBe stars. In order to place a lower limit on the disk accretion rate, we conservatively assume that 25% of the non-disk luminosity (460 $L_\odot$) is reprocessed by the disk; the disk then has an accretion luminosity of $L_D{}^{\mathrm{acc}} = (280 - 115) L_\odot = 165 L_\odot$.[5] This value of $L_D$ corresponds to a disk accretion rate of $\dot{M}_D \gtrsim 1 \times 10^{-5} M_\odot \,\mathrm{yr}^{-1}$ (eq. 3) which is consistent with the evidence for energetic mass outflow from 1548C27 and with the large disk accretion rates characteristic of Herbig AeBe stars.[6]

Applying the non-LTE model with these parameters, we obtain a good fit to the detailed properties of both spectral regions with Lorentzian line profiles and the parameters shown in Table 2. In each case, the synthesized spectrum for the 2-0 bandhead region is scaled by $\sim 95\%$ to match the continuum in both spectral regions (Fig. 6). In the 5-3 bandhead region, the true continuum is 0.246 Jy whereas the apparent continuum of 0.26 Jy (e.g., as measured at $2.382\mu$m ) is formed by the overlap of rotationally broadened lines. The "wiggles" in the emission redward of the peak in the 2-0 bandhead region arise from the "beating" of the (double-peaked) rotational broadening function against the rest wavelength distribution of the overtone lines. For systems with reduced projected rotational motion (e.g., 1548C27 and SVS-13), this effect appears close to the 2-0 bandhead where the spacing of lines is comparable to the separation of the horns of the rotational broadening function. The high resolution overtone spectroscopy by Carr & Tokunaga (1992) and Chandler et al. (1993, 1995) show the same effect.

---

[5] This probably underestimates the actual accretion luminosity because the disk albedo is probably nonzero and because the reprocessed radiation should not be counted separately from that of the primary source in computing the total system luminosity. Note, however, that the deduction of $L_{\mathrm{sys}}$ from observations depends on the viewing angle.

[6] Because the CO dissociation radius is larger than the stellar radius and the stellar contribution to the 2.3$\mu$m continuum is negligible, it is possible to find other stellar parameters which do not substantially alter the fit but correspond to different system luminosities. Note also that if the distance to 1548C27 is 0.9 kpc (Vilchez et al. 1989), the system luminosity is reduced by a factor of 7 and the disk accretion rate could be as low as $10^{-6} M_\odot \,\mathrm{yr}^{-1}$.

As shown in Figure 6, we are able to fit the 2-0 shoulder, peak, blue wing, and slope of the emission redward of the peak reasonably well with $\gamma$ in the range $1.0\,\mathrm{km\,s^{-1}}$ (Model G1_10) to $5.0\,\mathrm{km\,s^{-1}}$ (Model G5_9). The model "wiggles" would have reduced amplitudes if we convolved the synthetic spectra with the instrumental profile, but they would still appear too prominent when compared with the observations. This indicates perhaps a smoother inner cutoff to the CO abundance than given by the chemical and mechanical equilibria calculations of this paper. Apart from this problem, the relative heights of the emission at 2.3800 $\mu$m and 2.3815 $\mu$m are well fit by both models, as is the shape of the 5-3 bandhead at 2.3829 $\mu$m.

With $\gamma = 1\,\mathrm{km\,s^{-1}}$, there is a relatively narrow range of $T_\ell$ and $\sigma_\ell$ within which acceptable fits can be obtained. If the outer radius to the emission is roughly held fixed, much flatter $\sigma_\ell$ and steeper $T_\ell$ distributions would steepen the slope redward of the 2-0 bandhead and weaken the 2-0 wing relative to the fit in Model G1_10. This is because both a flatter $\sigma_\ell$ and a steeper $T_\ell$ reduce the emission from small radii (the latter by dissociation), weakening the blue wing and steepening the slope redward of the bandhead. Although equally good fits to the 2-0 bandhead are possible for somewhat flatter $T_\ell$ and $\sigma_\ell$ distributions, these parameters would also overemphasize the strength of the 2-0 P lines in the $v$=5-3 bandhead region. Temperature distributions as flat as $T_\ell \sim r^{-0.3}$ can be ruled out by the shape of the 2-0 bandhead alone. In this case, the flatter $T_\ell$ distribution reduces the emission from small radii by reducing the line source function, steepening the slope redward of the bandhead and weakening the blue wing. Intrinsically broader lines that are less optically thick allow acceptable fits within a larger range of $T_\ell$ and $\sigma_\ell$ distributions. Model G5_9 with $T_\ell = 9200(r/R_*)^{-0.55}$ and $\sigma_\ell = 620(r/R_*)^{-2}$ also produces a good fit. Column density distributions as flat as $r^{-1}$ can fit both bandhead regions reasonably well with $T_\ell \sim r^{-0.6}$.

### 5.4. Discussion of the Non-LTE Effect

In these models, the strength of the feature at 2.381 $\mu$m and the shape of the bandhead at 2.384 $\mu$m are sensitive to the departure of the level populations from LTE beyond the radius at which $T_\ell$ drops below 2000K. This effective outer radius to the emission occurs somewhat beyond the radius at which H associates to form $H_2$, since $H_2$ has a cross-section for collisionally exciting CO that is smaller than that of H by almost two orders of magnitude (see Appendix B). When non-LTE effects set in, the upper vibrational levels are depopulated first, followed by lower vibrational levels. This sequence produces effective outer radii that are larger for the lower overtone transitions. Radiative trapping plays an equally important role in determining the radii at which these effects occur because the line layer column densities required to fit the overtone emission correspond to very large optical depths in the fundamental transitions. The low escape probabilities associated with these optical depths result in significant radiative trapping of fundamental transition photons (cf. Scoville et al. 1980).

At these large optical depths, photons can only escape in the wings of the line, so the degree of radiative trapping depends significantly on the assumed behavior of the wings of the line profile function. For example, the escape probability for gaussian lines depends asymptotically on the line optical depth as $\sim \overline{\tau}_\nu^{-1} (\log \overline{\tau}_\nu)^{-1/2}$ whereas the escape probability for lines with Lorentzian or Voigt profiles depends asymptotically on the line optical depth as $\sim \bar{\tau}_\nu^{-1/2}$ (see Mihalas 1978). The shape of the line profile function also affects which transitions contribute most significantly to the escape of fundamental photons.

At the column densities required to fit the overtone emission, escapes from line profiles with weak line wings, such as gaussian line profiles, are difficult because they must be made far out in the wings. As a result, only the high-$J_\ell$ transitions have any substantial escape probability due to the reduced population of these levels and their consequently lower optical depths. At the same time, since the population of these levels is small, their contribution to the effective escape probability (eq. 10) is also small because the weighting factor ($f_{J'}$) is also reduced by the level population. This radiative trapping results in the extension of LTE level populations to radii significantly beyond the radius at which H associates to form $H_2$.

In comparison, the probability of escape from lines with Lorentzian profiles is large enough that the transitions from relatively well populated $J_\ell$ levels contribute significantly, resulting in substantially larger effective escape probabilities. In the fit shown in Figure 6a-b, which has a Lorentzian line profile with a width of $\gamma = 1.0$ km s$^{-1}$, the vibrational level populations reach half their LTE value at $\sim 1800$K ($v$=5) to $\sim 1400$K ($v$=1). Correspondingly, the strong transitions in the $v$=5-3 bandhead region have source functions that depart from LTE at $\sim 1800$K ($v$=5-3), $\sim 1750$K ($v$=4-2), and $\sim 1550$K ($v$=2-0), introducing effective outer radii to the CO emission at these temperatures. In this particular case, this departure from LTE only affects the spectral fit through its impact on the strength of the 2-0 P lines in the $v$=5-3 bandhead region. This is because only the 2-0 P lines still have a significant optical depth at the temperature where the departure from LTE is significant: at $\sim 1800$K the optical depth of a representative 5-3 line is $\tau(5$-$3) \simeq 0.2$, at $\sim 1750$K $\tau(4$-$2) \simeq 3$, and at $\sim 1550$K $\tau(2$-$0) \simeq 11$.

To illustrate the magnitude of the non-LTE effect on these lines, in Figure 7a-b we also show the spectral fit obtained by the (artificial) extension of LTE level populations to 600K ($r = 100 R_*$). The synthesized $v$=2-0 bandhead is somewhat stronger at the peak and the "wiggles" are filled in by emission from larger radii. The overall shape of the $v$=2-0 bandhead is altered only slightly, because the lines in the $v$=2-0 bandhead have optical depth $\tau \lesssim 1$ at the radii at which the non-LTE effect would have set in. In contrast, the overall shape of the spectrum in the $v$=5-3 bandhead region is altered significantly by the prominence of the 2-0 P lines which have $\tau > 1$ to $r > 40 R_*$ are consequently enhanced due to emission from larger radii. This sensitivity illustrates the utility of the 2-0 P lines as diagnostics of low temperature gas. To illustrate the effectiveness of this non-LTE effect in truncating the emission beyond a given radius, we show in Figure 7c-d the spectral fit obtained with non-LTE level populations, the parameters of Model G1_10, and an abrupt truncation of the emission at the radius at which $T = 1500$K.

## 6. Discussion and Summary

We have successfully modeled the CO overtone bandhead regions of the observed spectra of the YSOs WL 16 and 1548C27 in terms of emission from Keplerian disks. Our observations provide some of the most compelling evidence to date for the existence of disks around young stars. Through the detailed study of this emission, we are able to place useful constraints on the detailed physical properties of the inner disk region. These results also indicate the possibility of dynamical mass estimates of young stars through the study of the emission from their surrounding disks. In the case of WL 16, we are able to place a useful constraint on the stellar mass due to the relatively high inclination of the system and the availability of additional observational constraints on the distance and system luminosity. In the case of 1548C27, observations of the $v$=2-0 and $v$=5-3 bandhead regions provide a particularly potent combination of diagnostics which reveal the onset of non-LTE level populations. In our non-LTE model for the emission from this source, the CO emission arises from gas at temperatures between $\sim 5000 - 1500$K. The inner radius is due to the dissociation of CO; the outer radius is the result of the departure of the vibrational level populations from LTE due to radiative escapes from suprathermally broadened lines. The need for this outer radius is clearly indicated by the strength of the 2-0 P lines in the $v$=5-3 bandhead region.

Thermal Doppler line profiles appear unable to reproduce either the shape of the $v$=2-0 and $v$=5-3 bandhead regions of 1548C27 (§5.3) or the strength of the $v$=2-0 bandhead emission from WL 16 given our adopted extinction to the source (§3.2). In contrast, Lorentzian line profiles with widths of $\gamma = 0.5 - 5$ km s$^{-1}$ produce excellent fits to the overtone bandhead regions of both sources. These considerations argue against the presence of dust in the temperature inversion region of the disk atmosphere as the sole explanation for the effective outer radius to the overtone emission. In the absence of turbulence, dust is believed to settle very quickly from the surface layers of a circumstellar disk (Weidenschilling & Cuzzi 1993). In the presence of turbulence, which may mix dust into the surface layers, the non-LTE effect discussed in the text will come into play independent of any radiative transfer role for the dust.

The relatively large inner disk temperature inversion and the steep temperature and column density distributions required to fit the overtone emission restrict the possible physical explanations for the origin of the inversion. These include irradiation of the disk surface by the star and funnel flow accretion shock; turbulent heating in a wind-disk interface; and chromospheric heating that results from the vertical propagation of MDH waves into the upper disk atmosphere. Calvet et al. (1991) have used an approximate irradiation model to explore the ability of stars to heat their circumstellar disk surfaces and whether the CO overtone bandheads consequently appear in emission or absorption as a function of stellar spectral type and disk accretion rate. However, in terms of the detailed properties of the inner disk region, the model predicts flatter temperature and column density distributions as well as smaller temperature inversions than are implied by our observations of WL 16 and 1548C27. The steeper temperature and column density distributions required to fit the observations could result, instead, from the interaction of a stellar wind with the circumstellar disk: as the wind blows over the disk, the disk atmosphere is stirred up and heated by turbulent dissipation, forming an *aeolosphere* (cf. Carr et al. 1993). Suprathermal widths may also characterize lines formed in disk chromospheres. If accretion in the inner disk is the result of the Balbus-Hawley instability (Balbus & Hawley 1991), the magnetic fields that participate in this instability may also propagate MHD waves into the upper disk atmosphere, producing a turbulent temperature inversion layer.

If the turbulent line broadening resulting from either of the latter two possibilities is similar in nature to that observed in spectral line profiles of turbulent interstellar gas (Falgarone & Phillips, 1990), the Lorentzian line profiles adopted in our synthetic spectra may reasonably represent the expected non-gaussian line wings. We have examined and rejected the alternative possibility that the enhanced line widths are due to pressure broadening. According to standard estimates for this broadening mechanism, the gas pressure in the model layers are too low by several orders of magnitude to be responsible for the line widths. We conclude that the line-emitting region in these sources is likely characterized by velocity widths much larger than those appropriate for a quiescent layer heated by the external radiation field of the star. Since external irradiation will tend to stabilize disk atmospheres against convection, we suggest that these turbulent motions arise from perturbing influences such as stellar winds or vertical MHD wave propagation.

The stellar properties and disk mass accretion rates implied by our observations also bear on the origin of the inner disk temperature inversion. In the case of WL16, the implied disk accretion rate is low, $\dot{M}_D \lesssim 2 \times 10^{-7} M_\odot \, \text{yr}^{-1}$, independent of whether we model the stellar component as a G-star close to the stellar birthline, or as an A-star close to the main sequence. Given the low disk accretion rate, heating due to irradiation by the relatively early type star may then provide a simple explanation for the temperature inversion in the upper atmosphere of the surrounding disk. The low disk accretion rate is also consistent with the lack of evidence for energetic outflow activity from this source; at such low wind mass loss rates, aeolospheric or chromospheric effects may be confined to a modest amount of line broadening which contributes to the strength of the $v=2$-$0$ bandhead emission. However, the steep temperature and column density distributions derived for the line-emitting region may indicate that these effects are also responsible for additional non-negligible heating.

In contrast, our models for 1548C27 indicate a large disk accretion rate, $\dot{M}_D \sim 1 \times 10^{-5} \, M_\odot \, \text{yr}^{-1}$. Although this large mass accretion rate is consistent with both evidence for energetic mass outflow from 1548C27 and the disk accretion rates of other young stars with stellar parameters similar to those deduced for this object (Herbig AeBe stars), it also places 1548C27 in a parameter regime in which the models of Calvet et al. (1991) would nominally predict that the CO lines should be in absorption rather than in emission. This discrepancy may be reduced by considering the impact of the funnel flows that are likely to accompany these large disk accretion rates if intermediate mass stars possess magnetospheres (cf. Najita et al. 1995): the UV emission from the accretion hot spots at the base of the funnel flows is an energetic source of radiative heating. The possibility of a large contribution to the temperature inversion of the inner disk from funnel flow hot spots introduces the possibility of another source of energetic heating for the inner disk. As described by Shu et al. (1994), stellar magnetic fields strong enough to truncate the disk and to give rise to a funnel flow are also strong enough also to drive a magnetocentrifugal X-wind from the inner disk region. An aeolosphere may be created as the lowermost streamlines of this induced X-wind sweep over the inner disk region, producing the suprathermally broadened lines and steep temperature and column density distributions implied by our modeling of the overtone emission from this source. Similar line broadening and inner disk heating may arise from a disk chromosphere.

An improved understanding of the origin of the inner disk temperature inversion requires a detailed theoretical study of the temperatures and column densities that characterize aeolospheres and disk chromospheres. In addition to indicating the degree to which aeolospheres and chromospheres can heat the upper disk atmosphere, these studies may also indicate the observational signatures of aeolospheres and chromospheres. When combined with observa-tions of the type presented in this paper, these distinguishing characteristics could be used to address other astrophysical issues. For example, the observational identification of aeolospheres as the origin of the disk temperature inversion would favor a stellar origin for the wind over a disk origin, whereas the identification of chromospheres as the origin of the temperature inversion would support the Balbus-Hawley instability as the physical process that gives rise to accretion in disks.

It is a pleasure to thank N. Calvet for her insightful comments and advice. This work was supported in part by a grant from the NASA Origins of Solar Systems program (NAGW-3595). We also acknowledge support from NASA grant NAGW-630 (A. E. G.), NSF grant AST-9114940, and NASA grant NAGW-2278 (J. S. C. and A. T. T.). This work was carried out while J.S.C. held a Columbus Fellowship at The Ohio State University, Department of Astronomy.

# Appendix A. Radiative Rates

## Approximate A-values

The Einstein A-values for individual fundamental transitions from an upper level $(v',J')$ to a lower level $(v,J)$ can be obtained from

$$A_{v'J',vJ} = \frac{64\pi^4}{3h}\tilde{\nu}^3 \langle v'J'|\mu(x)|vJ\rangle^2 \frac{|m|}{2J'+1} \tag{A1}$$

given the transition frequency $\tilde{\nu}$ and the electric dipole moment function $\mu(x)$ as a function of the reduced internuclear distance $x$. The quantity $m = [J'(J'+1) - J(J+1)]/2$. For the purpose of calculating the vibrational level populations, we approximate the expectation of $\mu(x)$, $M_v^{v'} = \langle v'J'|\mu(x)|vJ\rangle$, with the rotationless transition dipole moment $M_v^{v'}(0)$:

$$A_{v'J',vJ} = \frac{64\pi^4}{3h}\tilde{\nu}^3 [M_v^{v'}(0)]^2 \frac{|m|}{2J'+1}. \tag{A2}$$

and use transition frequencies $\tilde{\nu}$ calculated using energy levels from Mantz et al. (1975) and values of $M_v^{v'}(0)$ from Goorvitch & Chackerian (1994).

Values of $A^*_{v'v}$ (cf. eq. 11) can be calculated by adding the rates for the R- and P-transitions associated with each upper level, weighting by the population of the upper level, and summing over all upper levels:

$$A^*_{v'v} \equiv \sum_{J'} f_{v'J'}(A_{v'J',vJ'-1} + A_{v'J',vJ'+1}).$$

For the purpose of calculating $A^*_{v'v}$, we have approximated the rotational level populations as

$$f_{v,J} = \frac{(2J+1)}{(T/B_v)}\exp(-J(J+1)B_v/T)$$

where the rotational constants $B_v$ are from Mantz et al. (1975). Since $A^*_{v'v}$ varies weakly with temperature over the range $T = 200 - 8000$K, we further approximate $A^*_{v'v}$ with its value at 2000K.

## Approximate Escape Probabilities

The escape probability of fundamental photons averaged over rotational levels (eq. 11) are approximated in the following way. The Einstein B-values for the fundamental transitions are related to the A-values by

$$B_{vJ,v'J'} = \frac{c^2}{2h\nu^3} A_{v'J',vJ} \frac{2J'+1}{2J+1}.$$

With the approximate form for $A_{v'J',vJ}$ given by equation (A2),

$$B_{vJ,v'J'} = b_{v'v}\frac{|m|}{2J+1} \tag{A3a}$$

where
$$b_{v'v} = (32\pi^4/3h^2c)[M_v^{v'}(0)]^2. \tag{A3b}$$

The sum of the B-values for the R- and P-transitions from the same lower level is independent of $J$

$$B_{vJ,v'J+1} + B_{vJ,v'J-1} = b_{v'v}.$$

Therefore, $b_{v'v}$ is the effective B-value relating the vibrational levels $v$ and $v'$

$$B_{vv'}^* \equiv b_{v'v}. \tag{A3c}$$

The corresponding optical depths are obtained from these upward B-values (eq. A3) using

$$\frac{\pi e^2}{m_e c} f_{\ell u} = \frac{h\nu}{4\pi} B_{vv'}^* \frac{|m|}{2J+1}$$

for the absorption oscillator strength $f_{\ell u}$ in the usual expression for the line optical depth. We also use the following simple approximation for the vibration-rotational energy levels in order to calculate transition frequencies and stimulated emission terms:

$$E_{vJ} = v\theta_0 + J(J+1)B_0,$$

where $\theta_0 = 3084$K and $B_0 = 2.768$K. Accordingly, for the rotational level populations, we adopt

$$f_J = \frac{(2J+1)}{(T/B_0)} \exp(-J(J+1)B_0/T). \tag{A4}$$

Given these optical depths, the effective escape probability $\beta_{v'v}^*$ (cf. eq. 11) is determined by averaging the escape probabilities for the R- and P-transitions associated with each upper level, weighting by the population of the upper level, and summing over all upper levels:

$$\beta_{v'v}^* \equiv \frac{1}{2} \sum_{J'} f_{J'} (\beta_{J',J'-1} + \beta_{J',J'+1})$$

where $\beta_{J',J'-1} = \beta(\tau_\nu^{J',J'-1})$, $\beta_{J',J'+1} = \beta(\tau_\nu^{J',J'+1})$, and the rotational level populations are as approximated in equation (A4)

## Appendix B. Collisional Excitation of CO Vibrational Levels

The excitation of the rotation-vibration bands of CO has been discussed earlier in a variety of astrophysical contexts, e.g., Thompson (1973; late-type stars); Scoville, Krotkov, & Wang (1980; young stellar objects); and Ayres & Wiedemann (1986; the sun). Our discussion focuses mainly on collisional excitation by atomic hydrogen, where there are real differences between authors, in contrast to collisions with electrons and hydrogen molecules. We emphasize new experimental data on the role of $\Delta v \geq 1$ transitions and the results of Elitzur (1983) based on surprisal theory.

Ayres & Wiedemann (1986) gave a critical discussion of existing experiments and theory on H + CO vibrational excitation. They used the Landau-Teller parameters determined in the shock-tube experiment of Glass & Kironde (1983, henceforth GK) and standard theory (Herzfeld & Litovitz 1959) to obtain the $\Delta v = 1 - 0$ de-excitation rate coefficient $k(1-0, \text{H})$. GK found that the product of the post-shock pressure $P$ and the measured relaxation time $\tau$ is essentially constant in the temperature range of the experiment $T = 1000 - 3000$ K,

$$P\tau = 1.8 \pm 0.4 \times 10^{-8} \text{atm} - \text{s}. \tag{B1}$$

Substituting this into the standard theory for the relaxation time

$$k(1-0) = k_\mathbf{B} T / P\tau \left(1 - e^{-\theta/T}\right)^{-1} \tag{B2}$$

we get a useful, direct representation of the GK data,

$$k(\text{1-0, H}) = 7.57 \times 10^{-15} \text{ cm}^3 \text{ sec}^{-1} \; T(1 - e^{-\theta_o/T})^{-1}. \tag{B3}$$

It must be emphasized that GK measured only *one* relaxation time and that they used the traditional method of analysis, based on the Landau-Teller theory, where the collision rate coefficients obey the rules for dipole matrix elements of a *simple harmonic oscillator* (the same as for radiation), i.e., there are no $\Delta v > 1$ transitions. The GK result for H + CO, i.e., equation B3, is 100-20 times larger the rate coefficient they measured for $H_2$ in the temperature range from 1000-3000 K.

At the time of the Ayres & Wiedemann (1986) paper, there was little experimental information on the validity of the harmonic oscillator selection rules, which forbid $\Delta v \geq 2$ transitions. Hooker & Millikan (1963) did study the problem for *rare-gas and molecule* collisions with CO with shock tube experiments at 2000 K. From the characteristics of the fundamental and first overtone time signals, they concluded that the CO population is largely built up by a sequence of $\Delta v = 1$ collisions, i.e., by "climbing the vibrational ladder". They estimated that the $\Delta v = 2$ collisions cannot occur as frequently as 1/10 as often as $\Delta v = 1$ collisions. In a high-pressure experiment on vibrational relaxation of *electronically excited* CO in the $A^1\Pi$ level, Fink & Gomes (1974) found that $\Delta v = 2, 3$ collisions were almost as important as $\Delta v = 1$ collisions. However, it is likely that physical processes other than those operative for the ground electronic level are at work in this experiment.

A new generation of molecular beam experiments is beginning to provide information on the dependence of the H + CO collisional excitation rate coefficients with selected final vibrational and rotational quantum numbers (e.g., Wight & Leone 1983, McBane et al. 1991). Although the experiments have only been done so far at energies higher than are relevant for astrophysical applications involving CO (1-3 eV), they do indicate that $\Delta v \geq 2$ transitions are significantly smaller than $\Delta v = 1$ transitions and that they decrease in importance as the energy is decreased. This is in conflict with Elitzur (1983), who correctly criticized the complete ignoring of $\Delta v \geq 2$ transitions, but proposed rate coefficients which we believe are too large, especially for transitions involving large changes in $v$. Elitzur's work is based on a statistical theory known as "surprisal theory" (e.g., Levine & Bernstein 1976). We suspect that the data underlying his application of this method is inappropriate to the H + CO system. i.e., they refer to *different* systems and situations, e.g., the one piece of evidence cited for CO involves rare gas atoms de-exciting an *excited electronic level* of CO (Fink & Gomes 1974)

In this paper, we adopt the following rate coefficients:

*Electron Collisions* - The rate coefficient for $T \leq 5000$ K that we derive from laboratory experiments (e.g., Morrison 1988) is

$$k(1-0,\mathrm{e}) \approx 1.24 \times 10^{-11}\,\mathrm{cm^3 s^{-1}}\,T^{1/2}. \tag{B4}$$

On this basis, we can ignore electronic collisions compared with neutrals collisions for $x_\mathrm{e} \leq 10^{-3}$, a restriction satisfied by the disk atmospheres and winds considered here.

*Molecular Hydrogen and Helium* - We follow previous studies and use the results of shock-tube experiments expressed in Landau-Teller form,

$$k(1-0,\mathrm{X}) = k_\mathrm{B} T\, e^{-(AT^{-1/3}-B)} \left(1 - e^{-\theta/T}\right)^{-1} \mathrm{atm-s}. \tag{B5}$$

The parameters given by Milliken & White (1963) are: X = He, $A = 99$ atm-s and $B = 20.4$ atm-s; X=$H_2$, $A = 68$ atm-s and $B = 19.1$ atm-s.

*Atomic Hydrogen* - In accord with the above discussion, we use equation B3.

The available information suggests that $\Delta v \geq 2$ transitions are relatively unimportant although quantitative data is not yet available for the temperature range of interest. We ignore these collisions and, again consistent with available experimental information, assume that all $\Delta v = 1$ transitions have the same rates. The rate coefficients summarized above indicate that for the disk layer analyzed in this paper, only atomic and molecular hydrogen collisions are important. Accordingly, in the level population calculation described in §4, the total vibrational collisional excitation rate is

$$nC^*_{v'v} = n_H\, k(\text{1-0, H}) + n_{H_2}\, k(\text{1-0, } H_2). \tag{B6}$$

This rate depends on the ratio $n_H/n_{H_2}$ which is assumed to be the chemical equilibrium value. Given the A-values discussed above, the corresponding critical densities for the vibrational levels are $n_H \sim 10^{12} - 10^{13}\,\mathrm{cm}^{-3}$ over the temperature range $T = 2000 - 5000$K. The critical densities increase toward both lower temperatures (due to reduced rate coefficients) and higher densities (which favor $H_2$ over H).

## Table 1. WL16 Model Parameters

| Model[†] | $R_*$ ($R_\odot$) | $T_{c0}$ (K) | $p_c$ | $T_{\ell 0}$ (K) | $p_\ell$ | $\sigma_{\ell 0}$ (g cm$^{-2}$) | $q_\ell$ | $\phi_\nu$ |
|---|---|---|---|---|---|---|---|---|
| L12 | 5.0 | 3000 | 0.75 | 6000 | 0.75 | 5000 | 4.5 | Thermal Doppler |
| L14 | 5.0 | 3000 | 0.75 | 6000 | 0.78 | 370 | 3.8 | Lorentzian ($\gamma = 0.5$) |
| L15 | 2.08 | 3000 | 0.75 | 5810 | 0.78 | 310 | 3.7 | Lorentzian ($\gamma = 0.5$) |
| U2 | 5.0 | 3000 | 0.75 | 5500 | 0.75 | 200 | 2.8 | Lorentzian ($\gamma = 0.5$) |
| V2 | 2.08 | 3000 | 0.75 | 5900 | 0.78 | 280 | 3.5 | Lorentzian ($\gamma = 0.5$) |

$M_* = 2.5 M_\odot$, $L_* = 22 L_\odot$, $d = 160\,\mathrm{pc}$, $A_K = 2.0$, $i = 60°$, $R_{\rm in} = 1 R_*$, $r_0 = 5 R_\odot$
[†] Models L12, L14, and L15 use LTE level populations. Models U2 and V2 use non-LTE level populations.

## Table 2. 1548c27 Model Parameters

| Model | $R_*$ ($R_\odot$) | $T_{c0}$ (K) | $p_c$ | $T_{\ell 0}$ (K) | $p_\ell$ | $\sigma_{\ell 0}$ (g cm$^{-2}$) | $q_\ell$ | $\phi_\nu$ |
|---|---|---|---|---|---|---|---|---|
| G1_10 | 5.9 | 9800 | 0.75 | 9700 | 0.6 | 7100 | 2.5 | Lorentzian ($\gamma = 1$) |
| G5_9 | 5.9 | 9800 | 0.75 | 13000 | 0.7 | 800 | 2.0 | Lorentzian ($\gamma = 5$) |

$M_* = 4.0 M_\odot$, $L_* = 200 L_\odot$, $d = 2400\,\mathrm{pc}$, $A_K = 0.62$, $i = 30°$, $R_{\rm in} = 1 R_*$, $r_0 = 5.9 R_\odot$

# Figure Captions

Figure 1. The $v$=2-0 bandhead emission from WL 16 which shows the characteristic shape of bandhead emission from a rotating disk: a blue wing, a shoulder, and an intensity peak redward of the bandhead. The spectral region covered by our observations includes the R39 - R62 lines.

Figure 2. Schematic depiction of the convolution of the double-horned rotational broadening function (*heavy line*) with the rest distribution of lines near the bandhead (*light line*). The rotational broadening function is the profile of an isolated line from an inclined Keplerian disk that has a monotonically decreasing intensity distribution between an inner and outer radius. The rest spectrum of lines near the bandhead includes line blending due to local line broadening sources and assumes LTE level populations at a characteristic inner disk temperature.

Figure 3. Model fits to the $v$=2-0 bandhead region of WL 16 (*solid line*) superposed on the observed spectrum (*histogram*): (a-c) LTE Models L12, L14, and L15; (d) Model L12 including the H$^-$ line layer continuum opacity (see text); (e,f) non-LTE Models U2 and V2. The difference between the upper and lower solid lines in each panel indicates the model continuum deficit or excess relative to the observed continuum level.

Figure 4. Relative LTE optical depths (see text) for representative 5-3 (*light solid line*), 4-2 (*dotted line*), 3-1 (*short dashed line*), 2-0 (*long dashed line*) lines in the 5-3 bandhead region compared with that for a line at the 2-0 bandhead (*heavy solid line*).

Figure 5. The $v$=2-0 and $v$=5-3 bandhead regions of the spectrum of 1548C27.

Figure 6. Non-LTE model fits to both the $v$=2-0 and $v$=5-3 bandhead regions of 1548C27 with (a,b) Model G1_10 ($\gamma = 1$ km s$^{-1}$) and (c,d) Model G5_9 ($\gamma = 5$ km s$^{-1}$). The $v$=2-0 bandhead region has been scaled by 95% to match the continuum in both spectral regions.

Figure 7. (a,b) The spectral fit obtained with the parameters of Model G1_10 and the (artificial) extension of LTE level populations down to 600 K ($r = 100R_*$). (c,d) The spectral fit obtained with the same model parameters, non-LTE level populations, and an abrupt truncation of the emission beyond the radius at which $T = 1500$ K. As in Figure 6, the $v$=2-0 bandhead region has been scaled by 95% to match the continuum in both spectral regions. When compared with the fits shown in Figure 6, these figures show that non-LTE level populations introduce a sharp outer radius to the emission, an effect that is easily diagnosed by fitting these bandhead regions.

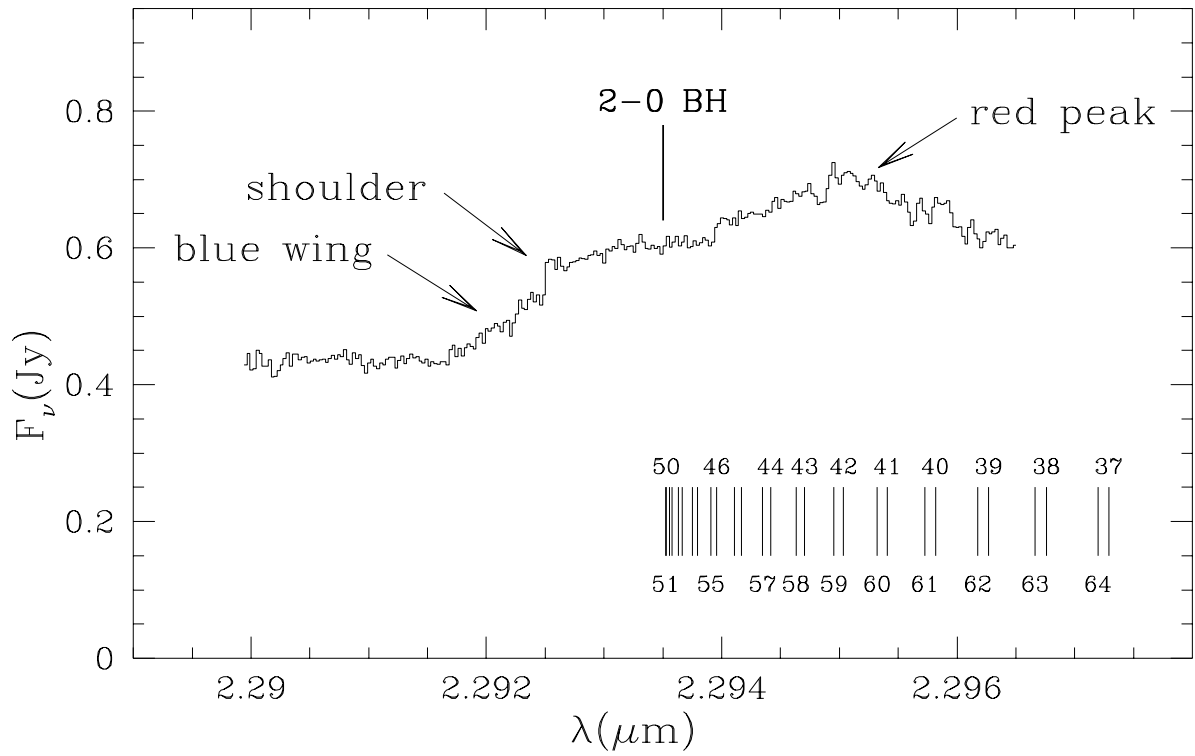

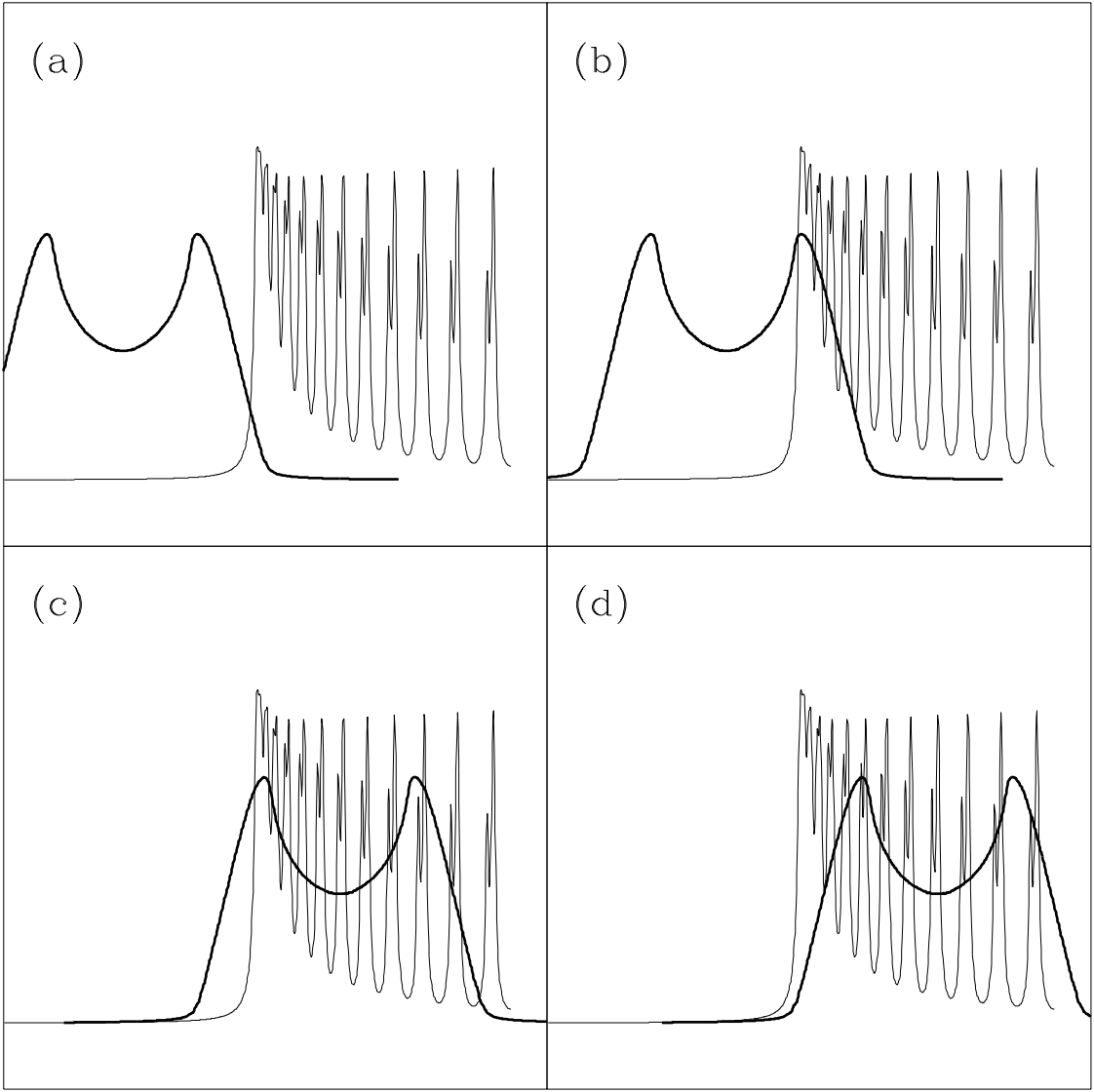

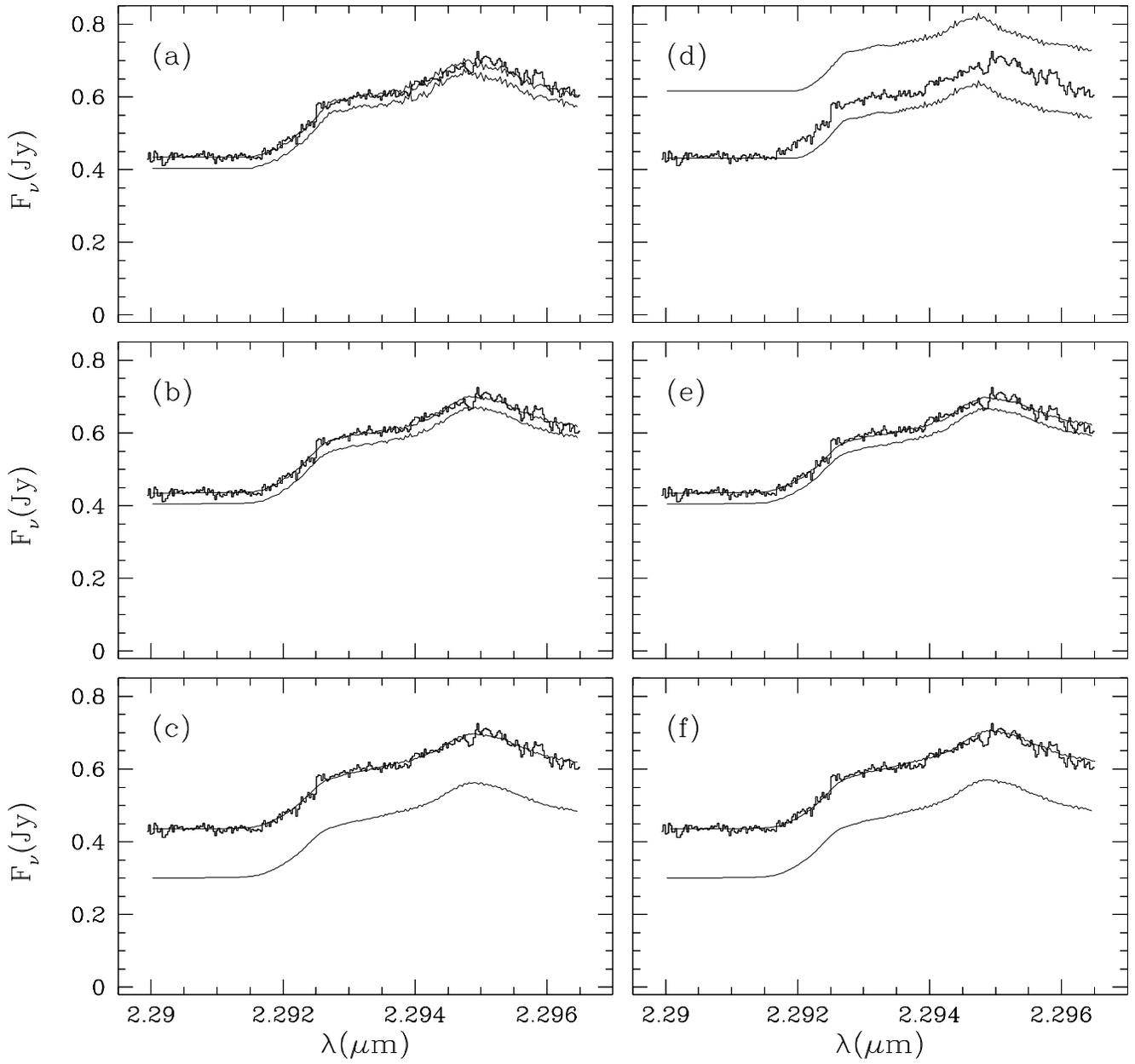

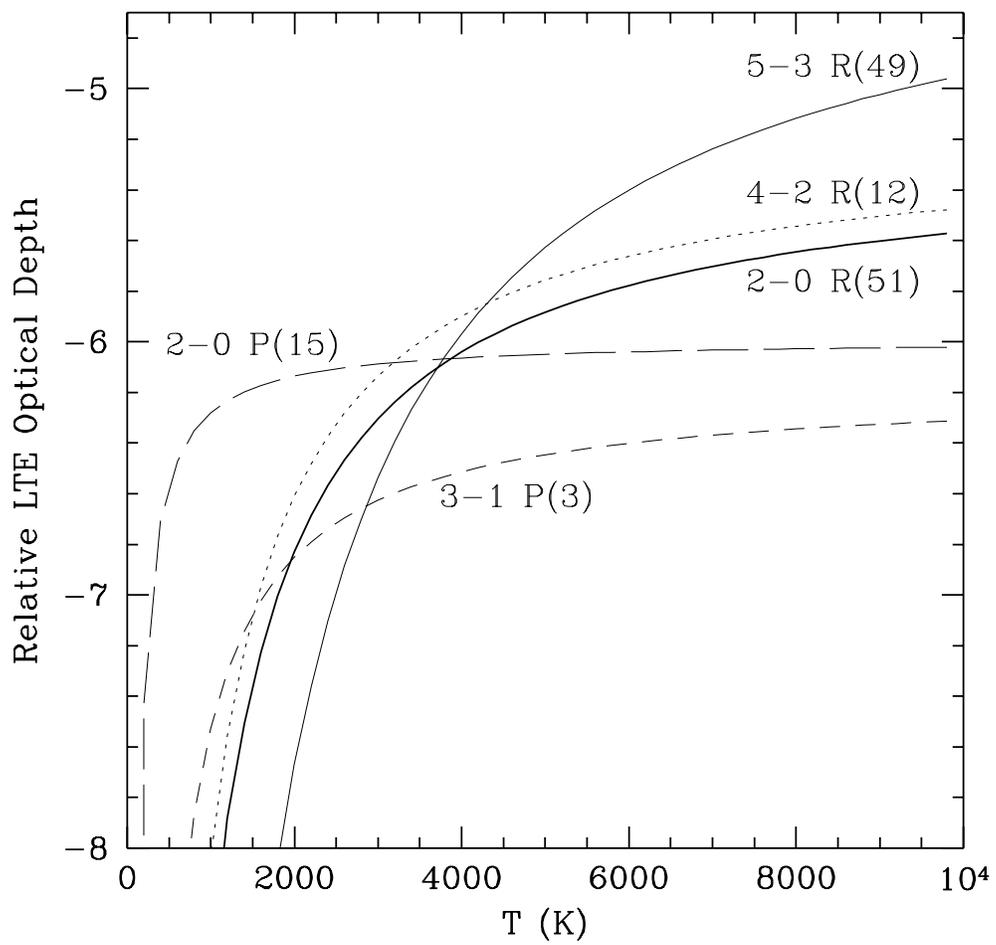

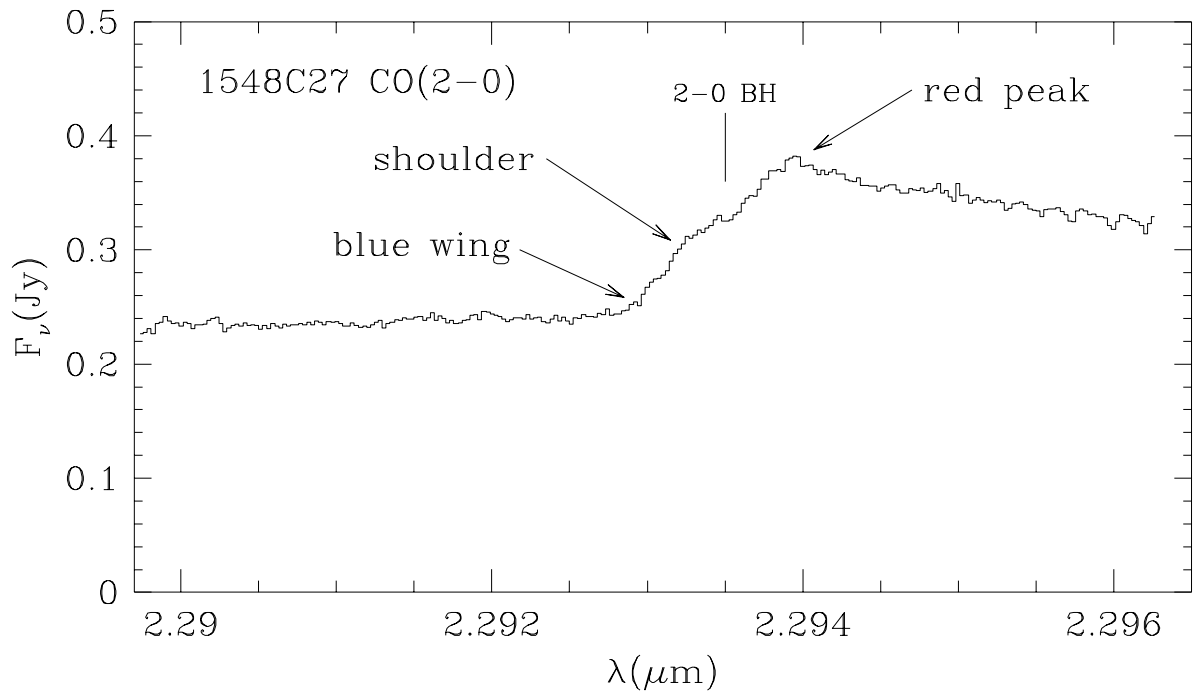
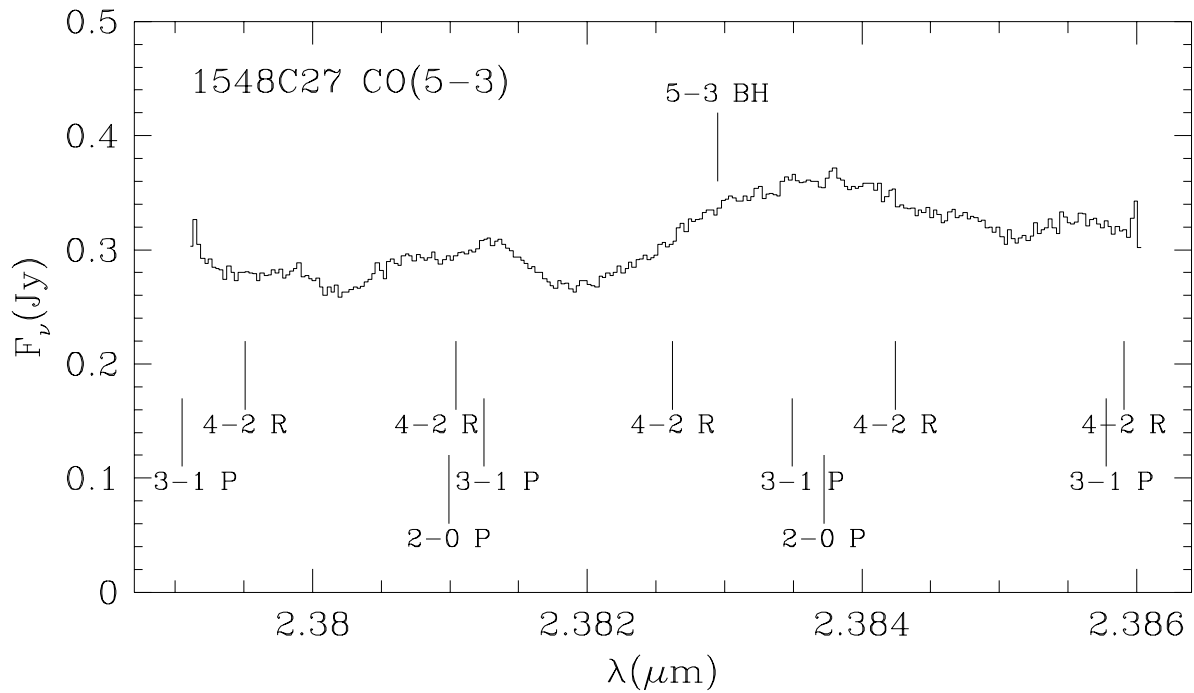

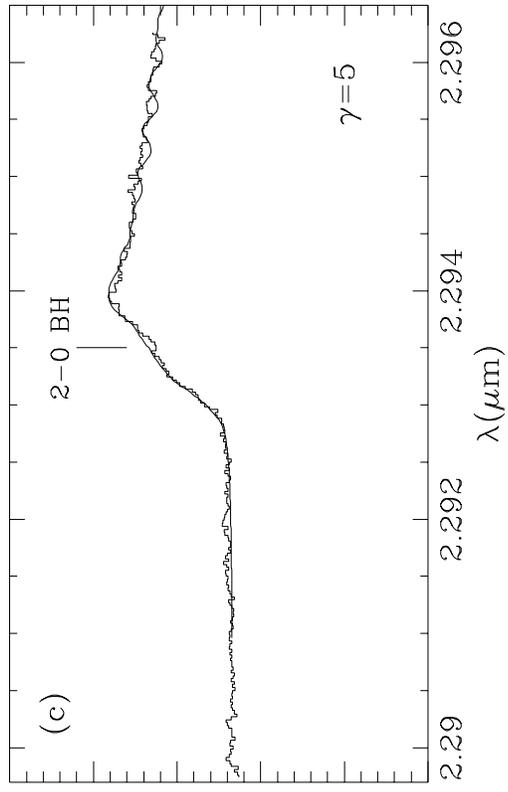

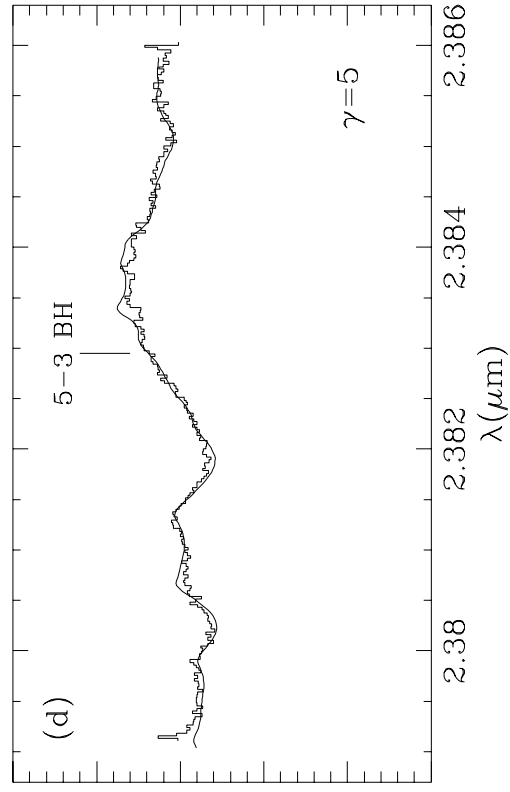

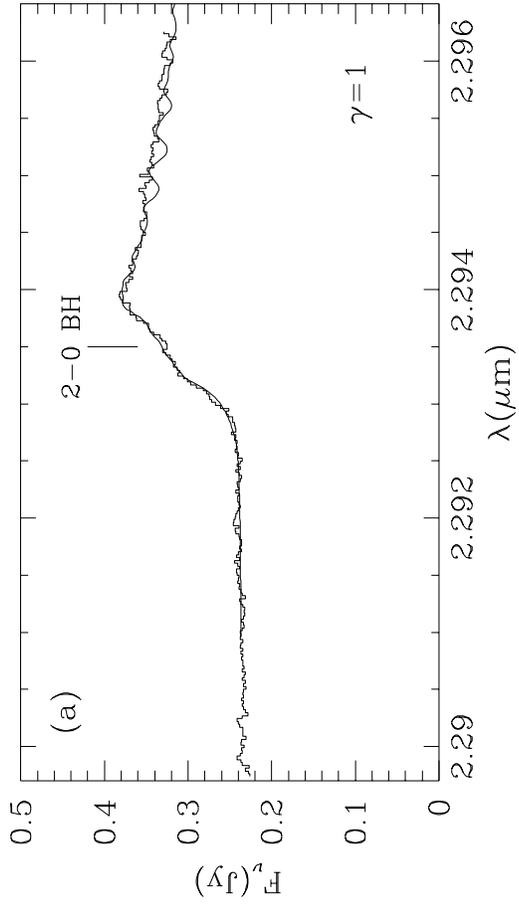

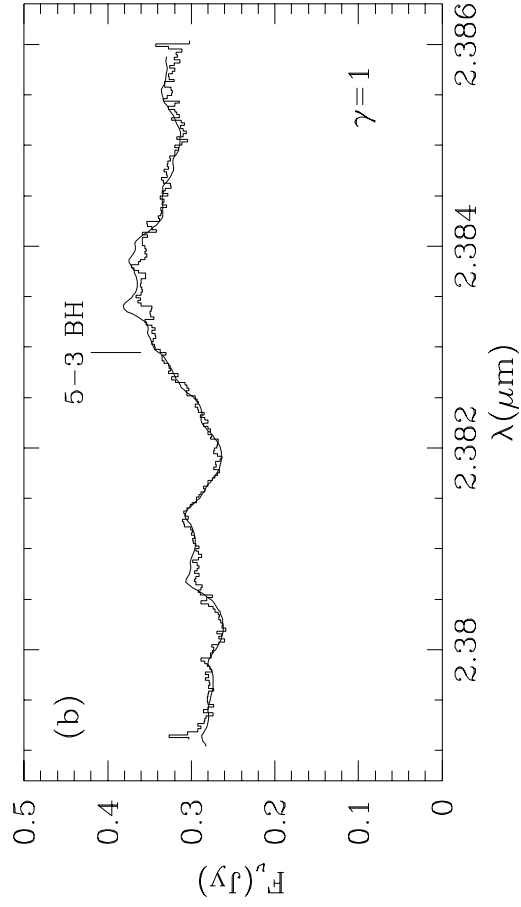

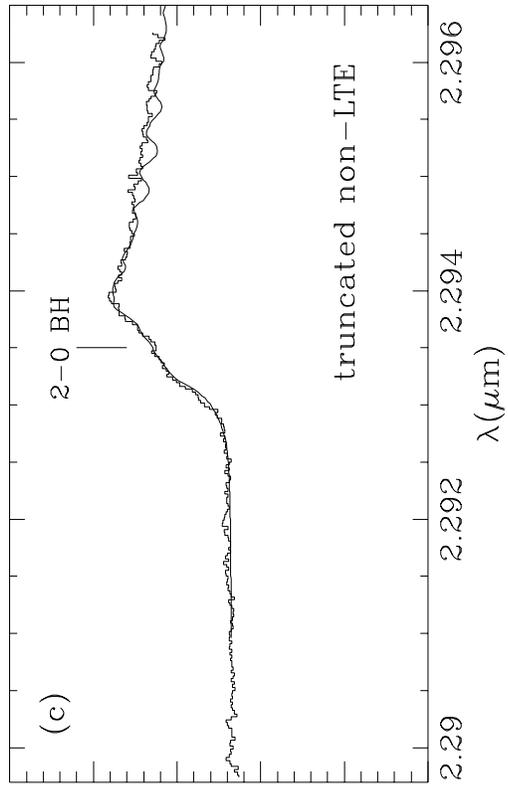
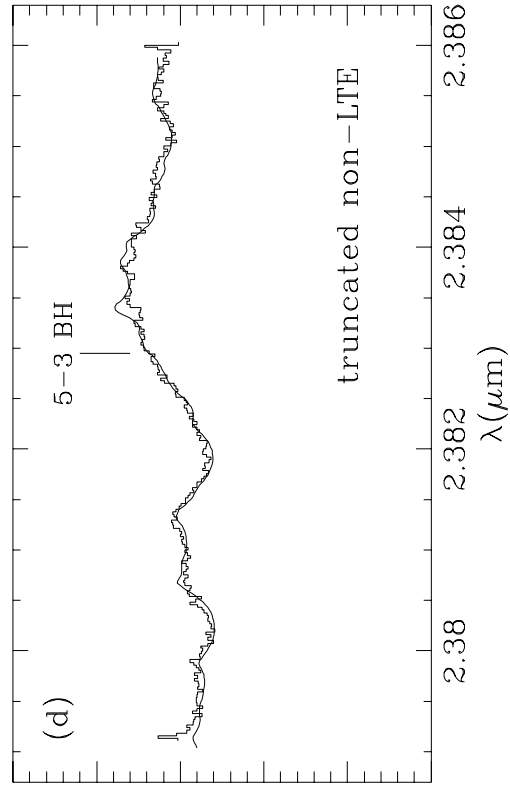
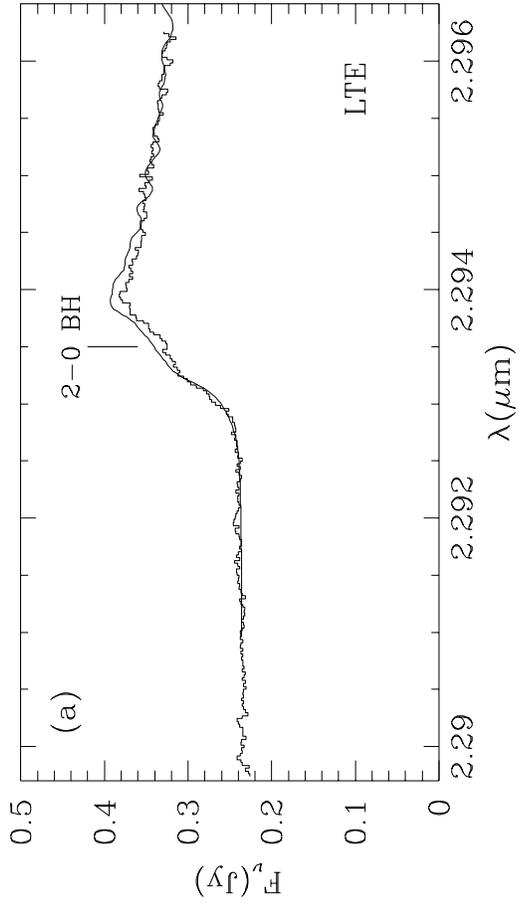
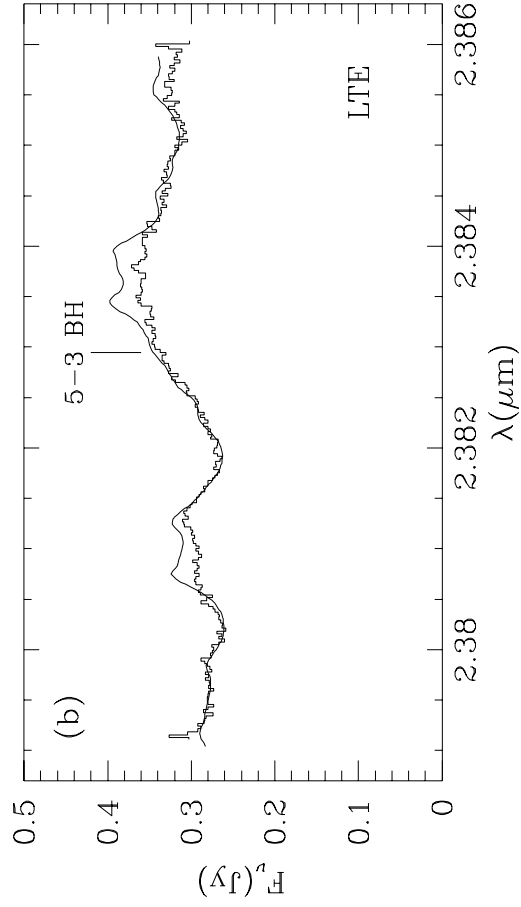